\documentclass[fleqn,usenatbib]{mnras}

\usepackage[T1]{fontenc}

\DeclareRobustCommand{\VAN}[3]{#2}
\let\VANthebibliography\thebibliography
\def\thebibliography{\DeclareRobustCommand{\VAN}[3]{##3}\VANthebibliography}

\usepackage{graphicx}	
\usepackage{amsmath}	
\usepackage{bm}
\usepackage[]{hyperref}
\usepackage{newtxtext,newtxmath}

\def\msun{{ ~M}_{\odot}}

\def\gpy{{\rm ~Gpc}^{-3} {\rm ~yr}^{-1}}
\def\kms{{\rm ~km} {\rm ~s}^{-1}}

\title[The role of SN convection for the lower mass gap in the isolated binary formation of GW sources]{The role of supernova convection for the lower mass gap in the isolated binary formation of gravitational wave sources.}

\author[A. Olejak et al.]{
Aleksandra Olejak,$^{1}$\thanks{E-mail: aolejak@camk.edu.pl}
Chris L. Fryer$^{2,3,4,5,6}$,
Krzysztof Belczynski $^{1}$
and Vishal Baibhav $^{7}$
\\
$^{1}$Nicolaus Copernicus Astronomical Center, Polish Academy of Sciences, 18 Bartycka Street, 00-716 Warsaw, Poland\\
$^{2}$Center for Theoretical Astrophysics, Los Alamos National Laboratory, Los Alamos, NM, 87545, USA\\
$^{3}$Computer, Computational, and Statistical Sciences Division, Los Alamos National Laboratory, Los Alamos, NM, 87545, USA \\
$^{4}$The University of Arizona, Tucson, AZ 85721, USA \\
$^{5}$Department of Physics and Astronomy, The University of New Mexico, Albuquerque, NM 87131, USA \\
$^{6}$The George Washington University, Washington, DC 20052, USA \\
$^{7}$ Center for Interdisciplinary Exploration and Research in Astrophysics (CIERA),
Northwestern University, 1800 Sherman Ave, Evanston, IL, 60201, USA\\
}

\date{Accepted XXX. Received YYY; in original form ZZZ}

\pubyear{2015}

\begin{document}
\label{firstpage}
\pagerange{\pageref{firstpage}--\pageref{lastpage}}
\maketitle

\begin{abstract}
Understanding astrophysical phenomena involving compact objects requires an insight about the engine behind core-collapse supernovae (SNe) and the fate of the stellar collapse of massive stars. In particular, this insight is crucial in developing an understanding of the origin and formation channels of the growing populations of detected black hole-black hole, black hole-neutron star and neutron star-neutron star mergers.\\ 
The timescale of convection growth may have a large effect on the strength of SN explosion and therefore also on the mass distribution of stellar remnants. We adopt new formulas for the relation between the pre-SN star properties and their remnants \citep{Fryer2022} and check how they impact population of double compact object (DCO) mergers formed via isolated binary evolution. 
The new formulas give one ability to test wide spectrum of assumptions on the convection growth time. In particular, different variants allow for a smooth transition between having a deep mass gap and a remnant mass distribution filled by massive neutron stars and low mass black holes. \\
We present distribution of masses, mass ratios and the local merger rate densities of DCO for different variants of new formulas and test them together with different approaches to other highly uncertain processes. We find that mass distribution of DCO mergers is sensitive to adopted assumption on SN convection growth timescale up to $m_1+m_2 \lesssim 35 M_{\odot} $. Between two extreme tested variants the probability of compact object formation within the mass gap may differ up to $\sim 2$ orders of magnitude. 
\end{abstract}




\section{Introduction}

Recent surveys provide more and more candidates and confirmed observations for black hole (BH) and neutron star (NS) systems in the Universe. Especially, increasing amount of gravitational wave (GW) detections significantly enriched the database with known compact objects opening new possibilities for studying and constraining their formation. So far LIGO/Virgo/KAGRA collaboration announced detection of GW signals from around $\sim 90$ sources ~\citep{LIGO2019a,LIGO2019b,Abbott2021a,Abbott2021b}. 
One of the most important questions in GW astrophysics is about the origin of compact binaries and the formation scenario of the detected DCO mergers. The popular formation scenarios in the literature include: the isolated binary evolution ~\citep{Bond1984b,Tutukov1993,Lipunov1997,Belczynski2010a,2014A&A...564A.134M,Kinugawa2014,Hartwig2016,deMink2016,Mandel2016a,Marchant2016,Eldridge2016,Woosley2016,Heuvel2017,Stevenson2017,Hainich2018,Marchant2018,Spera2019,Neijssel2019,Buisson2020,Bavera2021,Qin2021,vanSon2021,Olejak2021b}, the dense stellar system dynamical channel~\citep{PortegiesZwart2000,Miller2002a,Gultekin2004,Gultekin2006,OLeary2007,Sadowski2008,Downing2010,Antonini2012,Benacquista2013,Mennekens2014,Bae2014,Chatterjee2016,Mapelli2016,Hurley2016,Rodriguez2016a,VanLandingham2016,Askar2017,ArcaSedda2017,Samsing2018,Morawski2018,Belczynski2018,Banerjee2018,DiCarlo2019,Zevin2019,Rodriguez2018a,Perna2019,Kremer2020a}, isolated multiple (triple, quadruple) systems \citep{Toonen2016,Antonini2017b,Silsbee2017,Arca-Sedda2018,LiuLai2018,
Fragione19,Gomez2021,Stegmann2022}, mergers of binaries in galactic nuclei \citep{Antonini2012,Hamers2018,Hoang2018,Fragione2019b,Tagawa2020,Tagawa2021}.
Progress in understanding the origin of compact binaries requires having a reliable relation between properties of pre-SN star and its remnant mass (NS, BH or no stellar remnant). Conversely, in order to have a reliable distribution of remnant masses, scientists need to constrain existing SN models by comparing results of population synthesis models with the detected and observed population of compact objects. Several recent studies applied population synthesis calculations to predict mass distributions of Galactic or cosmological DCO population adopting different SN prescriptions \citep{Zevin2020,Shao2021, Dabrowny2021,Bavera2022}. Such comparisons will become much more reliable in the near future as the predicted number of DCO mergers is going to increase significantly after O4 run \citep[e.g.][]{Magee2022} and later, thanks to the development of next-generation ground-based gravitational wave detectors \citep{Borhanian2022}. \\
There are several approaches trying to mimic the true SN engine in order to predict the final fate of massive stars and estimate the remnant masses. However, the final faith of progenitors with their initial masses in the range $20-40 \msun$, so eventual lower mass gap fillers, are especially challenging as the outcome may drastically differ depending on the accumulated explosion energy \citep{Fryer2012,Liu2021}.\\ Due to complexity of the problem and emergence of numerical viscosity in 3-dimensional modeling, some finds more reliable using 1-dimensional codes with artificially altered energy deposition into pre-SN star in order to get an explosion. One approach is to alter the neutrino luminosity or absorption to increase the energy deposited~\citep{2006ApJ...637..415F,2010A&A...517A..80F,2012ApJ...757...69U,2015ApJ...806..275P,2016AJ....152...41P,2016ApJ...818..124E}. Another approach is to implement 1-dimensional mixing models to drive explosions~\citep{2018ApJ...856...63F,2020ApJ...890..127C}. 
In this study we assume that convection dominates the matter motion above the proto neutron star \citep{herant94,2003ApJ...584..971B,Fryer2007,2015ApJ...808L..42M,2018SSRv..214...33B,Fields2021} and therefor is crucial for SN engine modelling. We implement and test new results for different 1-dimensional mixing prescriptions of \cite{Fryer2022}. The timescale of convection growth significantly affect the course of eventual SN explosion and therefore also the final distribution of compact object remnants \citep{Fryer2007}.\\
The new formulas give one ability to test a wide spectrum of assumptions on the convection growth time. In particular, different variants of the formulas allow for a smooth transition between having a deep lower mass gap and a remnant mass distribution filled by massive NSs and low mass BHs. The dearth of compact objects with the masses in the range $\sim 3-5 \msun$ among observed population of X-ray binary systems \citep{Bailyn_1998,Farr_2011,Ozel_2010} has led to idea of potential mass gap between the maximum mass of a NS and the minimum mass of a BH. Also the latest discoveries and analyzes of \cite{LIGOfullO32021, LIGOfullO3population2021} points to the possible existence of a NS/BH mass gap. On the other hand, some recent estimates for isolated compact objects detected via microlensing give provide candidates for NS/BH mass in a gap range \citep{Mroz_mass_gap_2021,mic_2022,Mic_mass_gap_2022}. Due to a relatively small sample of known compact objects and usually significant error bars on their mass estimates, it is not yet clear if the mass gap is real phenomenon. The possible physical mechanism responsible for formation of the gap, the rapid SN timescale of the convection growth, was suggested by \cite{Fryer2012,Belczynski2012}. This work is an update and continuation of those studies.
In this study we implement new formulas for SN remnant masses into the {\tt StarTrack} population synthesis code and test how different variants of mixing (corresponding to different convection growth timescales) change distribution of DCO mergers masses, mass ratios and local merger rate density. Section \S \ref{sec: Method} explains the used method, including a brief description of the {\tt StarTrack} code. It contains our main adopted physical assumptions, especially the description of new adopted prescriptions for remnant masses. In Section \S \ref{sec: Single_Evol} we present relation between properties of pre-SN stars (NS or BH progenitor) with its new remnant mass in single star evolution together with comparison with previously used models. In Section \S \ref{sec: Binary_Evol} we present results for isolated binary evolution for different types of DCO mergers. Section \S \ref{sec: Conclusions} includes summary of the results and conclusions. 
Furthermore, Appendix \ref{sec:Mcrit} demonstrates how different assumption on the mass threshold for BH formation (instead of NS) influences the lower mass gap. In Appendix \ref{sec:redshift_BHBH_rates} we attach redshift evolution of BH-BH merger rate densities for two formation channels with and without common envelope phase (CE). Appendix \ref{sec:Stoachsticity} includes the effect of eventual stochasticity in pre-SN stellar structure on the mass distribution of DCO mergers. In Appendix \ref{sec:MOBSEcomparios} we briefly compare our new SN prescriptions with other parameteraized prescription for stellar remnant masses proposed by recent study \cite{Dabrowny2021}. 

\section{Method} \label{sec: Method}

\subsection{StarTrack code} \label{sec: StarTrack}

In this study we generate a population of cosmological DCO mergers using {\tt StarTrack} population synthesis code \citep{2002ApJ...572..407B,2008ApJS..174..223B}. The version of the code used in this paper is the same as the one described in Sec. 2 of \cite{Olejak2021a} with two important modifications: the first is the new remnant mass formulas (see Sec. \ref{sec: New_formulas}) and the second is different approach to pair-instability supernovae (PPSN)/pair-instability supernovae (PSN). In this study we test two different prescriptions for PSN. In the first approach we adopt strong PPSN/PSN which limits mass of BHs to $\sim 45 \msun$ as adopted in \cite{Belczynski2016c}. The second approach is revised prescription from \cite{Belczynski2020PSN} in which star experience disruption in PSN if the final mass of its helium core is in the range $90\msun <M_{\rm He} <175 \msun$. The revised PSN model does not include any mass loss in PPSN and allows for formation of BH with mass up to $90 \msun$. In the strong PPSN/PSN model we use initial mass function (IMF) with the maximum mass for a star limited to $150 \msun$, while we extend it to $200 \msun$ in the revised PSN model, similarly as done for \cite{Belczynski2020PSN}.

We use a model of star formation history and metallicity distribution in the Universe \citep{MadauFragos2017} described in \cite{2020A&A...636A.104B}. 
Adopted procedures for accretion onto a compact object during stable RLOF and from stellar winds are based on the analytic approximations \citep{King2001}, implemented as in \cite{Mondal2020}. For non-compact accretors, we assume a 50\% non-conservative 
RLOF \citep{Meurus1989,Vinciguerra2020} with a fraction of the donor mass ($1-f_{\rm a}$) lost from the system together with corresponding part of the donor and orbital angular momentum \citep[see \S 3.4 of][]{2008ApJS..174..223B}. In the current version of {\tt StartTrack} code we limit mass accretion rate onto non-compact accretor to the Eddington limit. We use the formula:
\begin{equation}\label{eg:Eddington limit}
\dot{M}_{\rm Edd}= \frac{4 \pi c R}{\eta k_{th}} . 
\end{equation}
The adopted limit is derived under the
assumptions that accretion goes to a star radius $R$ and there are no outflows/jets so $\eta=1$. Here $c$ is the speed of light and $k_{th}$ is Thomson scattering opacity. Eddington limit for non-compact accretor may be exceeded in our simulations in case of thermal-timescale mass transfer from massive, rapidly expanding donor on its Hertzsprung gap. We adopt 5\% Bondi-Hoyle rate accretion onto the compact object during the CE phase \citep{Ricker&Taam, MacLeod2015a,MacLeod2017} and standard {\tt StarTrack} physical value for the envelope ejection efficiency $\alpha_{\rm CE}$ = 1.0.
For the stellar wind mass loss we use formulas based on theoretical predictions of radiation driven mass loss \citep{Vink2001} with the inclusion of Luminous Blue Variable (LBV) mass loss \citep{Belczynski2010a}. We adopt Maxwellian distribution natal kicks with $\sigma=265\kms$ \citep{Hobbs2005} lowered by fallback \citep{Fryer2012} at NS and BH formation.

While presenting results for binary evolution we distinguish between two prescriptions for stability of RLOF: the standard CE development criteria described in Sec. 5.2 of \cite{2008ApJS..174..223B} and the revised mass transfer stability criteria based on results of \cite{Pavlovskii2017}, implemented and tested in \cite{Olejak2021a}. The revised RLOF stability criteria applies only to systems with massive donors (mainly BH progenitors) with their initial masses over $20 \msun$. It allows for CE development under much more restricted conditions than the standard {\tt StarTrack criteria}, taking account the system mass ratio at RLOF onset, the radius of the donor and metallicity. For details see Section 3.1 or the revised stability diagrams plotted in Fig. 2 and 3 of \cite{Olejak2021a}. Simplifying, the typical critical mass ratio (of donor to accretor) for a system to develop CE instead of stable RLOF is around 4-5 for the revised criteria while it is rather 2-3 for the standard {\tt StarTrack} criteria. 

\subsection{New formulas for remnant masses} \label{sec: New_formulas}

The main modification in the {\tt StarTrack} code tested in this paper is implementation of new formulas for remnant masses (NSs and BHs) given by \cite{Fryer2022}. In their study, \cite{Fryer2022} apply sub-grid mixing algorithms to the profile of post-bounce pre-SN cores in order to follow the convection growth above proto-neutron stars, estimate explosion energies and the final remnant masses. In particular, the new formulas represent analytical fits to different solutions obtained for a set of one dimensional core-collapse models based on mixing-length theory and a Reynolds-averaged Navier Stokes approach \citep{Livescu2009}. For more details see Sections 2 and 3 of \cite{Fryer2022}.

\cite{Fryer2022} predict the final masses of formed NS and BH by capturing how differences in convection growth timescales associated with variations in mixing affects the final stellar remnant masses. Increasing the mixing length tends to accelerate the growth of convection making explosion more likely. Such a trend is commonly observed in 1-dimensional simulations results. Based on those results, \citep{Fryer2012} obtained simple mathematical fits which incorporate the pre-SN stellar structure, to estimate the remnant mass for a given progenitor.

In contrast to the previously used rapid and delayed SN models given by formulas 5 and 6 of \cite{Fryer2012} that may be treated as two extremes, new formula gives one ability to test remnant masses for wide spectrum of assumptions on convection growth timescales.
The formula \ref{eq:mrem} allows to calculate remnant masses assuming smooth relation with pre-SN carbon-oxygen core mass $M_{\rm CO}$ and varying different mixing efficiency ($f_{\rm mix}$) set by the convection growth time.

\begin{align} 
    M_{\rm rem} = & 1.2+0.05 f_{\rm mix}+0.01 (M_{\rm CO}/f_{\rm mix})^2 + \nonumber \\ 
    & e^{f_{\rm mix}(M_{\rm CO}-M_{\rm crit})}.
    \label{eq:mrem}
\end{align} 
Where $M_{\rm rem}$ -- a remnant mass (NS or BH); $f_{\rm mix}$ -- convection mixing parameter which takes value in the range 0.5-4.0;  $M_{\rm CO}$ -- mass of the carbon-oxygen core of the pre-SN star in $\msun$ unit (in {\tt StarTrack} it is the value at the end of star core helium burning phase), $M_{\rm crit}=5.75M_\odot$ -- the assumed critical mass of carbon oxygen core for a BH formation (the switch from NS to BH formation).
Note that the mass of remnant is calculated using the Equation \ref{eq:mrem} till some $M_{\rm CO}$ value, which depends on the steepness of the exponent (so the adopted $f_{\rm mix}$ value). If $M_{\rm rem}$ from Eq. \ref{eq:mrem} exceeds the value of the total mass of pre-SN star ($M_{\rm pre-SN}$), then we assume the direct collapse of the star to a BH with only mass loss in form of neutrinos ($1 \%$ of pre-SN mass):  
\begin{equation}
    M_{\rm rem} = min(M_{\rm{eq.}~\ref{eq:mrem}},M_{\rm pre-SN}).
\label{eq:rem}
\end{equation}

Convection mixing parameter $f_{\rm mix}$ (Eq. \ref{eq:mrem}) is inversely proportional to convection growth time. Therefore,
$f_{\rm mix}=4.0$ corresponds to rapid growth of the convection $<$10ms that then develops an explosion in the first 100ms (depending on mass) while $f_{\rm mix}=$0.5 corresponds to a growth time closer to 100ms where the explosion can take up to 1s.
New formula with adopted value $f_{\rm mix}\approx 0.5$  result in shallow or no mass gap similarly to previous delayed SN model \citep{Fryer2012} while high value of $f_{\rm mix}\approx 4.0$ result in deep mass gap, similarly to previous rapid SN model. Applied formulas are for non-rapidly rotating progenitors.

\subsection{Detection weighted calculations} \label{subsec: Detections}
The distributions of the source properties observed by gravitational-wave detectors are biased due to the selection effects.
We account for such detection biases when we compare mass-ratio distributions of the synthetic population of DCO mergers (Sec. \ref{subsec:mass_ratio}) with the population detected so far by The LIGO-Virgo-KAGRA Collaboration (LVK).
We assume that a given merger is detectable if it has the signal-to-noise ratio (SNR)$>8$. The SNR for each merger can be expressed as $\rho=\rho_0 w>8$, where $\rho_0$ is the SNR assuming the binary is optimally oriented and located in the sky, and $0\leq w\leq 1$ is the projection factor that depends on the binary's sky position and orientation. We calculate the SNRs using the waveform approximant IMRPhenomD (\cite{Khan:2015jqa}), assuming LIGO mid-high sensitivity (corresponding to O3 observing run). 

For each binary within the detector's horizon (i.e. $\rho_0>8$), we find the probability that it will be detected,
\begin{equation}
p_{\rm det}= P(8/\rho_0) 
\end{equation}
where $P(w)$ is the cumulative probability distribution function of $w$ (\cite{Finn:1992xs}). Each merger in our population is weighted by $p_{\rm det}$ to account for detector effects.

\section{Single star evolution} \label{sec: Single_Evol}

In this section we show results for new remnant mass formulas only for a single star evolution. We present relations between progenitor star mass at its Zero Age Main Sequence (ZAMS) versus: the final remnant mass (NS or BH), the total mass of the pre-SN star, masses of pre-SN carbon oxygen (CO) and helium core (He) for different metallicities. For comparison, besides the two extreme examples of new remnant mass formulas \citep{Fryer2022}, we provide similar results for the previous {\tt StarTrack} models: the rapid and delayed SN model \citep{Fryer2012}. 
In Sec. \ref{sec:MassGap} we present how different assumption on convection growth timescale, corresponding to different $f_{\rm mix}$ variants (see Eq. \ref{eq:mrem}), impacts the depth and width of the lower mass gap.

\subsection{Remnant mass vs its stellar progenitor} \label{sec:ZAMS}

In Figures ~\ref{fig:standard_engnes} - ~\ref{fig:revised_engines}, we present the final stellar remnant mass (NS or BH) as a function of its progenitor's ZAMS mass for four models of SN: the two previous {\tt StarTrack} models, so-called the delayed and rapid SN models (Fig. \ref{fig:standard_engnes}) and two models chosen from the spectrum provided by \cite{Fryer2022} (Fig. \ref{fig:revised_engines}). The two examples of new SN models correspond to extreme variants for the mixing parameter value: $f_{\rm mix}$=0.5 and $f_{\rm mix}$=4.0 (see \S ~\ref{sec: Method}). Every plot includes three panels which correspond to different stellar metallicities:
at the top 1\% $Z_{\odot}$, in the middle 10\% $Z_{\odot}$ and at the bottom 100\% $Z_{\odot}$ ($Z_{\odot} = 0.02$). For every model we show two variants for PSN limit (see Sec. \S \ref{sec: Method}). Note that the ZAMS mass range in the figures is extended up to $M_{\rm ZAMS}=300 \msun$ while in our cosmological simulations for binary systems in next sections we limit possible initial mass of the star to $150 \msun$ for strong PPSN/PSN model and to $200 \msun$ for the revised PSN model (see Sec. \S \ref{sec: Method}).

\subsubsection{Differences between previous and new SN models} \label{subsec:low_mass}

\begin{figure*} 
	\includegraphics[width=8.85 cm]{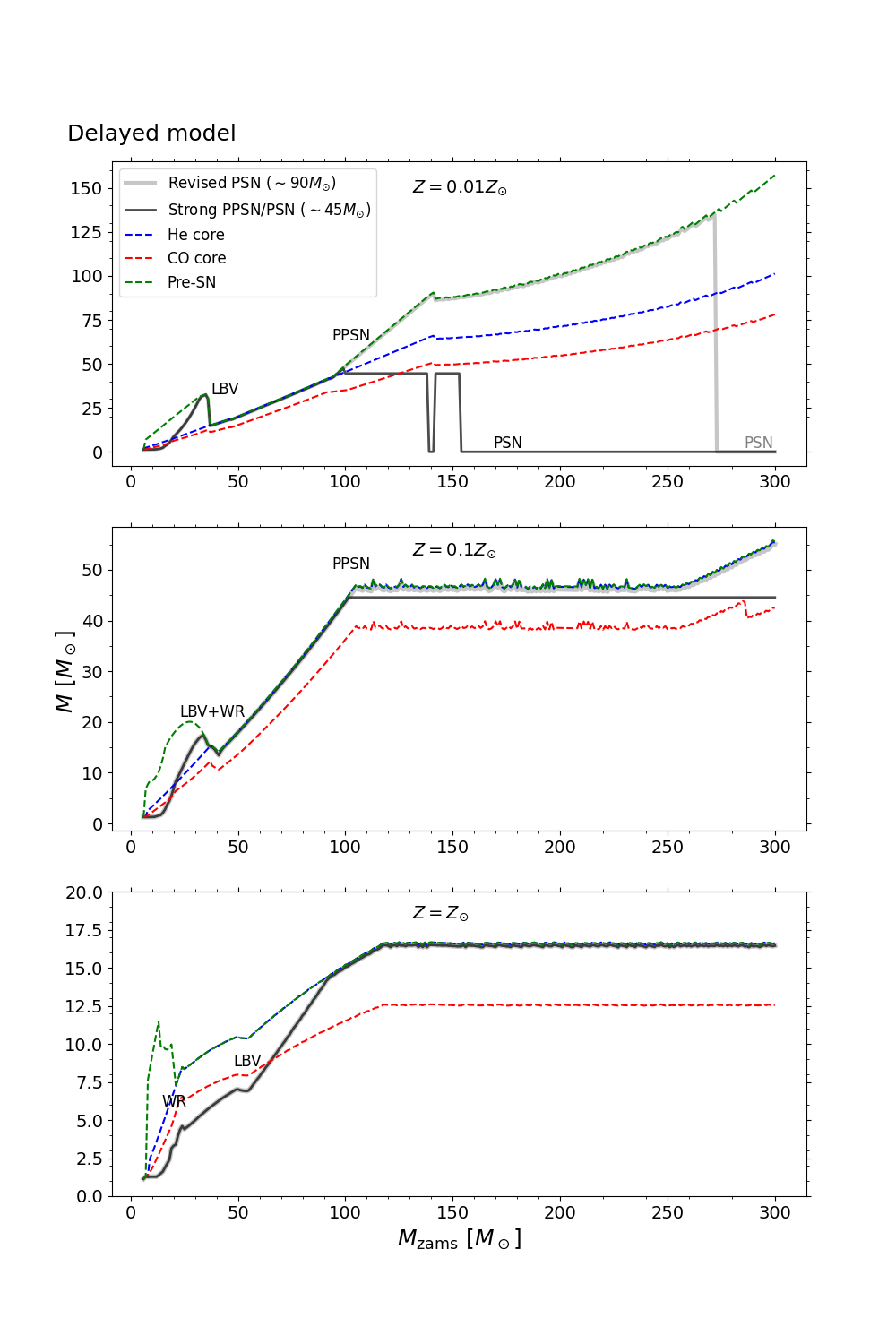}
	\includegraphics[width=8.85 cm]{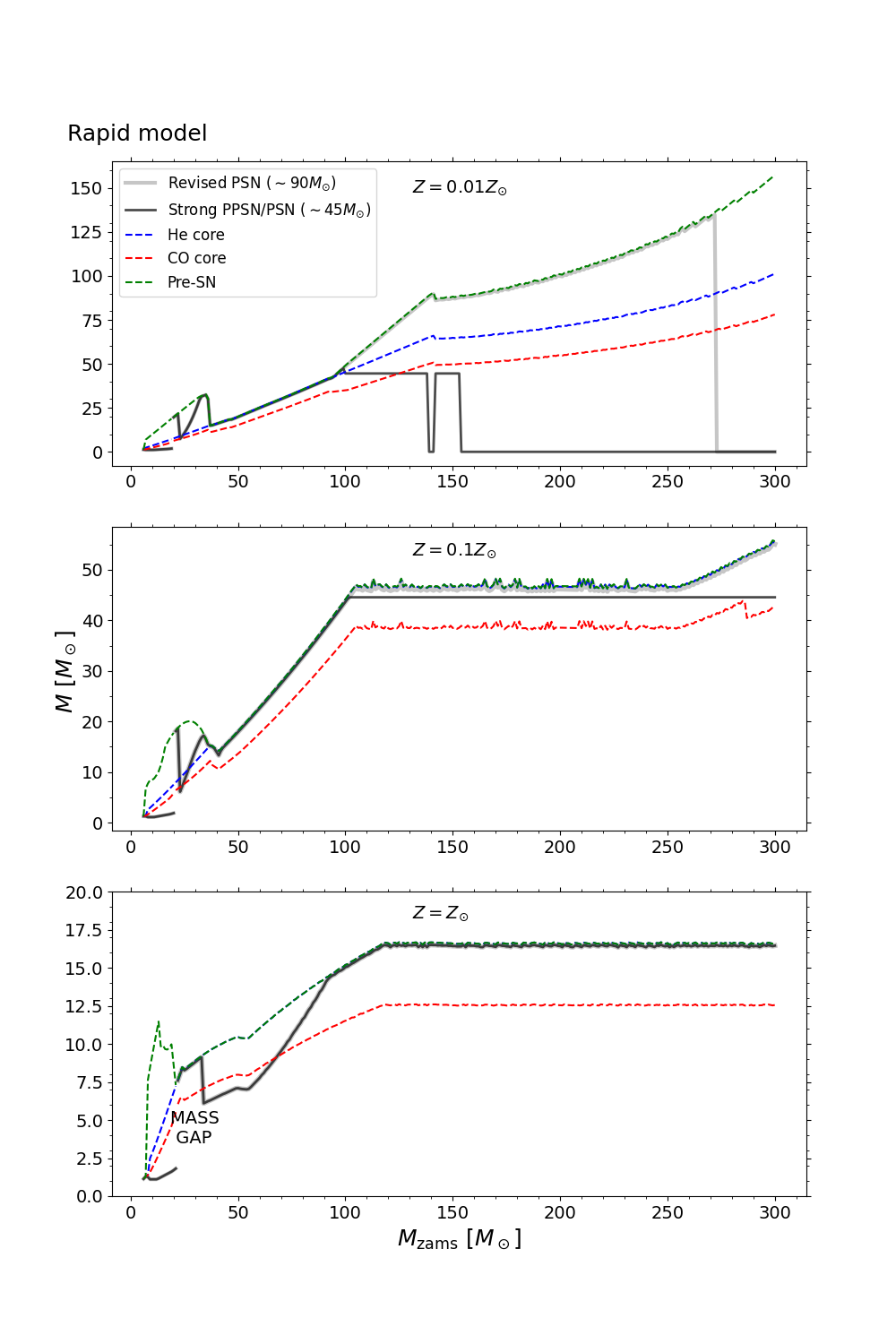}
    \caption{\textbf{Left:} Remnant mass (NS or BH) and pre-SN star properties as a function of progenitor ZAMS mass for the delayed SN model \citep{Fryer2012}. Top panel -- $1\% Z_{\odot}$;  middle panel -- $10\% Z_{\odot}$; bottom panel -- $100\% Z_{\odot}$ ($Z_{\odot}=0.02$;). Black line -- mass of the remnant (BH or NS) for strong PPSN/PSN model; grey line - mass of the remnant (BH or NS) for revised PSN model; blue dashed line -- mass of the helium core of pre-SN star; red dashed line - mass of the carbon-oxygen core of pre-SN star; green dashed line -- total mass of the pre-SN star. 
    \textbf{Right:} Same results for the rapid SN model \citep{Fryer2012}.}
    \label{fig:standard_engnes}
\end{figure*}

\begin{figure*} 
	\includegraphics[width=8.85 cm]{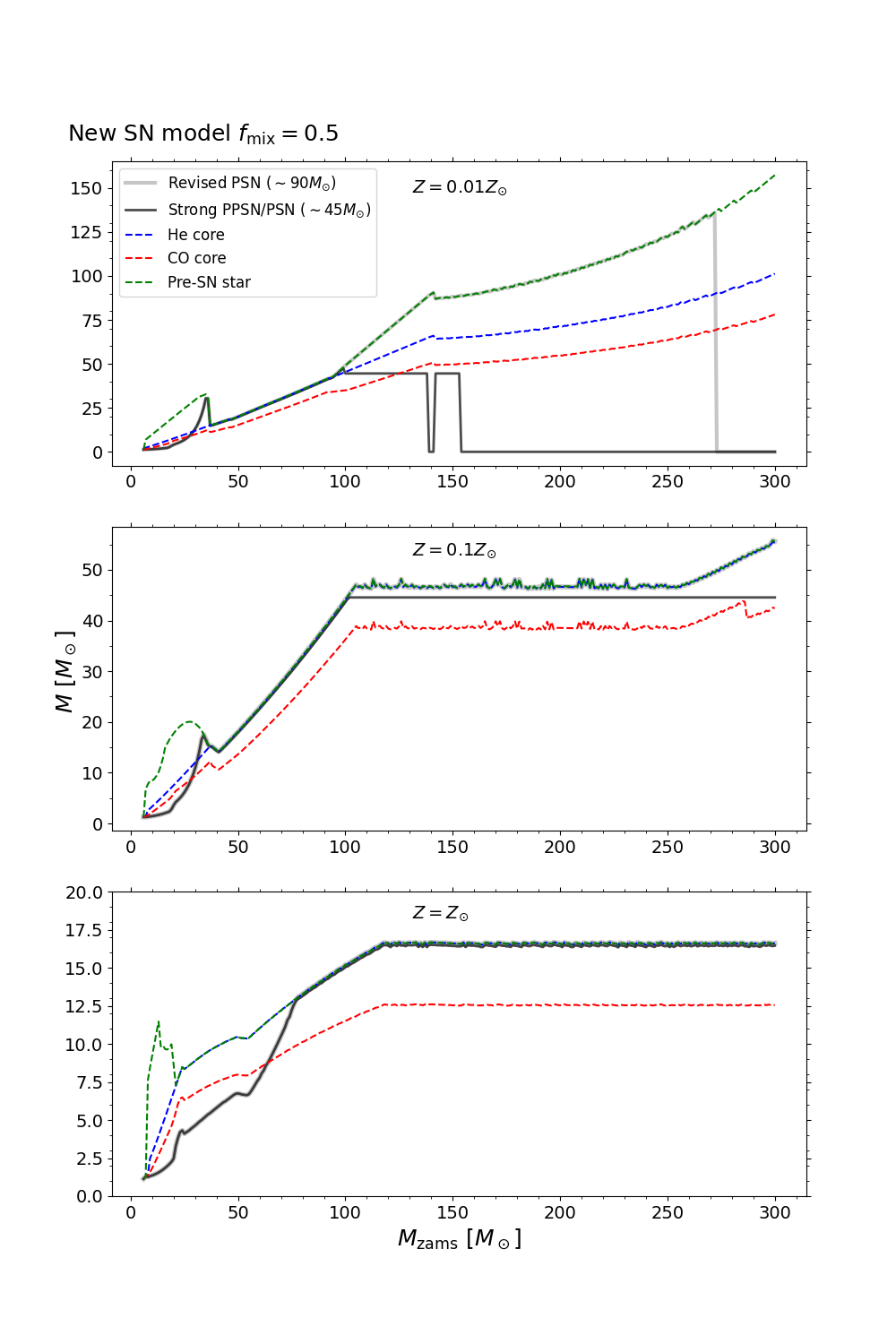}
	\includegraphics[width=8.85 cm]{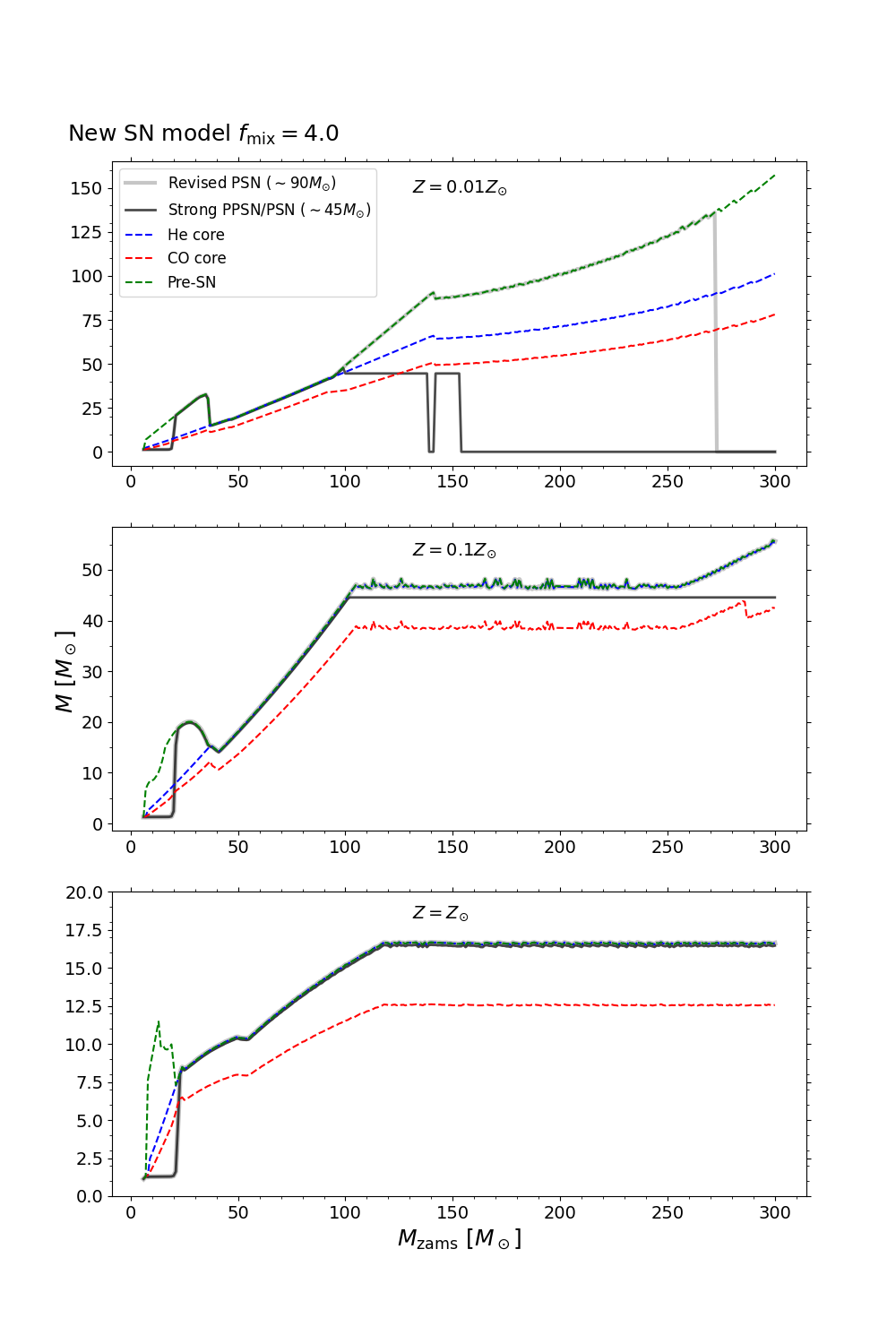}
    \caption{\textbf{Left:} Remnant mass (NS or BH) and pre-SN star properties as a function of its progenitor ZAMS mass for the new SN model with $f_{\rm mix}=0.5$ \citep{Fryer2022} Top panel -- $1\% Z_{\odot}$;  middle panel -- $10\% Z_{\odot}$; bottom panel -- $100\% Z_{\odot}$ ($Z_{\odot}=0.02$;). Black line -- mass of the remnant (BH or NS) for strong PPSN/PSN model; grey line - mass of the remnant (BH or NS) for revised PSN model; blue dashed line -- mass of the helium core of pre-SN star; red dashed line - mass of the carbon-oxygen core of pre-SN star; green dashed line -- total mass of the pre-SN star. 
    \textbf{Right:} Same results for the new SN model with $f_{\rm mix}=4.0$ \citep{Fryer2022}. Note that new SN models with low values of $f_{\rm mix}\approx 0.5$ (in Eq. \ref{eq:mrem}) result in shallow or no mass gap (similarly to old delayed SN model) while high values such as $f_{\rm mix}\approx 4.0$ result in deep mass gap (similarly to previous rapid SN model).}
    \label{fig:revised_engines}
\end{figure*}

The adopted core collapse SN model plays a role up to some progenitor ZAMS mass threshold above which the fates of all collapsing stars end in the direct collapse to a BH with minimal mass ejection. After exceeding this threshold, which depends on metallicity, the relation between $M_{\rm ZAMS}$ and remnant mass of most massive stars (the right side of the plots) looks the same for all SN models (Fig. ~\ref{fig:standard_engnes} - ~\ref{fig:revised_engines}).

The main difference between the previous and new adopted formulas for remnant masses is that the new ones allow for probing convection growth timescale on broad spectrum, which also means probing the depth and width of lower mass gap, instead of testing only two extreme cases: delayed and rapid SN models. The extreme cases of the new formulas: $f_{\rm mix}=0.5$ and $f_{\rm mix}=4.0$, may be defined as substitutes of the previous delayed and rapid SN engines respectively. Those models are, however, not exact equivalents, especially the previous rapid and new $f_{\rm mix}=4.0$ differ by few important features. In order to compare the major features of SN models, we will divide them into two categories: the slow convection growth models: previous delayed SN and new $f_{\rm mix}=0.5$ and the fast convection growth models: previous rapid SN and new $f_{\rm mix}=4.0$. 

The major difference between the slow and fast convection growth SN models is the limit on initial mass $M_{\rm ZAMS}$ threshold for a direct collapse into a BH (Fig. ~\ref{fig:standard_engnes} - ~\ref{fig:revised_engines}). For the fast SN models, the threshold for $M_{\rm ZAMS}$ separating progenitors ending evolution in a direct collapse instead of successful SN explosion is noticeably lower than that of the slow SN models. This is because, when the convection growth timescale is fast (fast SN models) for progenitors with masses $M_{\rm ZAMS}\geq =20 \msun$, convection is strongest when the material is still able to prevent an explosion. In contrast, for the same progenitors but once convection growth in slow timescale (slow SN models), the peak in the convection occurs when the infalling ram pressure is weaker, allowing an explosion, see \cite{Fryer2012,Belczynski2012}.   The threshold for direct collapse in fast SN models is around $M_{\rm ZAMS}=20 \msun$ while for slow SN models it is around $M_{\rm ZAMS}=35 \msun$ for lower metallicities (1\% and 10\% $Z_{\odot}$) and around $M_{\rm ZAMS}=80 \msun$ for 100\% $Z_{\odot}$ as for for such a high metallicity the stellar winds removes significant part of star's mass. 

SN models with a slow convection growth timescale, so the previous delayed and new model with $f_{\rm mix}=0.5$, give very similar results, with a small difference that the new model produce slightly lower SN remnant masses than the previous one. But between the two models with fast convection growth, previous rapid and the new model with $f_{\rm mix}=4.0$, there are two important differences. The low threshold for a direct collapse to a BH (around $M_{\rm ZAMS}=20 \msun$) results in a mass gap betwen NS and BH masses for both fast SN models. In previous model, this mass gap was totally empty - zero compact objects produced in the mass range: $\sim 2-5\msun$. Our new SN models, even the most rapid convection growth, with $f_{\rm mix}=4.0$, allow for formation of some compact objects within this range. However, the depth and width of the mass gap between NS and BH masses (fraction of compact objects with masses $\sim 2-5\msun$) changes with the adopted value of $f_{\rm mix}$ (see Sec. \ref{sec:MassGap} and Fig. \ref{fig:MassGap}). The new SN model with $f_{\rm mix}=4.0$, which corresponds to the most rapid convection growth, allows for a slight filling of the mass gap and is less extreme than the previous rapid model \citep{Fryer2012}. It is a result of implementation of a better understanding of the growth time and narrowing the range of assumptions on mixing.

The second difference between the previous rapid and new model with $f_{\rm mix}=4.0$ is the lack of an additional dip for remnant masses after exceeding the direct collapse threshold ($\sim M_{\rm ZAMS}=20 \msun$) in the new SN model. The origin of the dip in the previous rapid model was motivated by the feature observed in detailed evolutionary set of models \citep{Woosley2002} for high metallicity progenitors $Z=Z_{\odot}$ \citep{Fryer2012,Belczynski2012}. In those models, stars with their initial masses above the $M_{\rm ZAMS} \geq 20\msun$ due to increased mass loss of the outer layers in stellar winds have their cores structure modified. As a consequence the density of heavy oxygen and silicon layers may be decreased so the energy required to eject the outer part of the star is reduced. This, in turn leads to resuming successful SN explosions until the initial mass of progenitors reach other limits at which the final star is so massive that gravity leads to direct collapse to a BH. The position of this additional dip as well as its extent depends on the used detailed code and is sensitive to choice of many input physical parameters such as mixing or mass-loss prescriptions. For example in the versions of code used for studies by \cite{Fryer2022} the dip is less pronounced than it was for \cite{Fryer2012}. Including dip modelling into new formulas would make them more complex and would require considering results for different star metallicity.  This study does not include this effect.

\subsubsection{Progenitor masses $M_{\rm ZAMS} < 100\msun$} \label{subsec:middle_mass}

There are few features in the relation between the ZAMS mass of the progenitor and the final remnant mass of the NS/BH which do not originate from the adopted SN model but from other physical processes. For example, LBV and Wolf-Rayet star (WR) winds play a role after exceeding a certain mass thresholds for progenitor initial mass. Those thresholds, however, strongly depend on metallicity. In Table \ref{tab:LBV_winds} we provide values of $M_{\rm ZAMS}$ for which massive stars in our simulations become subject of increased mass loss due in the LBV or WR winds for three metallicities: 1\% $Z_{\odot}$, 10\% $Z_{\odot}$ and 100\% $Z_{\odot}$. We also marked the origin of features: LBV, WR, PPSN and PSN in Figure \ref{fig:standard_engnes}

\begin{table}
	\centering
	\caption{Mass threshold for progenitors ZAMS mass to enter the regime of additional mass loss in LBV/WR winds for three different metallicities. }
	\begin{tabular}{lccr} 
		\hline
		 & $M_{{\rm ZAMS}, Z_\odot}$ & $M_{{\rm ZAMS},0.1Z_\odot}$ & $M_{{\rm ZAMS},0.01Z_\odot}$ \\
		\hline
		LBV & $50 \msun$ & $33 \msun$ & $32 \msun$\\
		WR & $24 \msun$ & $38 \msun$ & $37 \msun$\\
		\hline
	\end{tabular}
    \label{tab:LBV_winds}
\end{table}

In all SN models (Fig. ~\ref{fig:standard_engnes} - ~\ref{fig:revised_engines}), we observe a clear dip in the final remnant mass at lower metallicities (1\% $Z_{\odot}$ and 10\% $Z_{\odot}$) for progenitors with initial masses $M_{\rm ZAMS} \gtrsim 32 \msun$ due to entering the effects LBV winds. The luminosity of such massive stars exceed Humphreys–Davidson limit (\citep{Humphreys1994}) and stars are subject of significant additional mass loss in LBV stellar winds of order of $10^{-4} \msun {\rm yr}^{-1}$\citep{Belczynski2010}. Massive stars with high metallicities (like 100\% $Z_{\odot}$) are also subject of LBV winds in our simulations. However, for high metallicties stars lose more mass in stellar winds during the the early evolutionary phases (main sequence). It removes significant fraction of star's outer layers and shifts the threshold for LBV winds to larger progenitor masses $M_{\rm ZAMS}\gtrsim 50 \msun$.

Another important threshold for $M_{\rm ZAMS}$ corresponds to WR-star winds \citep{Hamann1998}. The minimum mass of stars to be a subject of WR winds, similar to LBV winds, depends on the metallicity. Metallicity also significantly impacts the duration of the WR phase (in our simulations the stripped helium core phase) and how much mass is lost in WR winds. In the case of WR winds, in contrast to LBV winds, the higher the metallicity, the lower the  mass $M_{\rm ZAMS}$ threshold for WR winds. Rich in metals, massive stars are subject of significant mass loss in stellar winds during their earlier stage of evolution that may allow them to loose quickly hydrogen envelopes. Poor in metals, massive stars may also lose significant part of their outer layers in stellar winds. However, it usually happens in the latter part of the evolution and therefore the duration of the WR phase (and its mass loss) is also shorter. As an example, the same progenitor star with $M_{\rm ZAMS}=40\msun$ at solar metallicity $Z=Z_{\odot}$ will lose $\sim 4\msun$ during its WR phase, at $Z=0.1 Z_{\odot}$ around $\sim 1\msun$ while at $Z=0.01 Z_{\odot}$ only $\sim 0.1\msun$.

For stars with 100\% $Z_{\odot}$, the $M_{\rm ZAMS}$ threshold for WR winds starts for progenitors with $M_{\rm ZAMS}\sim 24 \msun$ and is clearly visible in Figures \ref{fig:standard_engnes} and \ref{fig:revised_engines}. In the case of lower metallicities: 1\% and 10\% $Z_{\odot}$, the threshold at which WR winds activates is shifted to  $M_{\rm ZAMS}\sim 37 \msun$.  However, note that for 1\% $Z_{\odot}$ due to relatively short WR phase the feature is negligible.

\subsubsection{Progenitor masses $M_{\rm ZAMS} \gtrsim 100\msun$} \label{subsec:large_mass}
Most massive stars $M_{\rm ZAMS} \gtrsim 100\msun$ may experience significant mass loss in PPSN or complete disruption in the violent PSN explosion initiated by the creation of the electron-positron pairs which reduce the radial pressure in their cores \citep{Woosley2007, Woosley2017}. However due to several uncertainties associated with physical processes in massive stellar cores, such as the rates for $^{12}$C({\ensuremath{\alpha}}, {\ensuremath{\gamma}})$^{16}$O reaction, and a lack of strong observational evidence \citep{LIGOfullO3population2021}, it is currently not possible to determine the exact mass range stars are subjects of potential PPSN or PSN \citep{Farmer2020,2021Costa, Woosley2021}. 
In this study we test two different variants for the PSN limit. In our strong model \citep{Belczynski2016c}, the maximum mass of the BH is limited to $\sim 45 \msun$. Stars with final helium cores $M_{\rm He}>45 \msun$ are subject of PPSN while stars with  $M_{\rm He}>65 \msun$ are completely disrupted in a PSN explosion. The second variant is a revised PSN model by \cite{Belczynski2020PSN}. In the revised PSN model, we do not include PPSN mechanism while the limit for total star disruption in a PSN explosion is shifted to stars with final helium core masses $M_{\rm He}>90 \msun$.  The total disruption of the star in a PSN explosion for both PSN models happens only for the lowest tested metallicity and the effect is visible only on the top panels (for 1\% $Z_{\odot}$) of Figures ~\ref{fig:standard_engnes} and ~\ref{fig:revised_engines}. 
In the strong PPSN/PSN model, stars with progenitor masses $M_{\rm ZAMS} \gtrsim 100 \msun$ experience PPSN which decreases the final mass of BH to $45 \msun$. The non-linear, complex relation of stellar winds with the stellar mass and luminosity influences the final mass of the He core and makes it a non-monotonic function of $M_{\rm ZAMS}$. As an effect, the mass of He cores first exceeds $65 \msun$ (limit for complete disruption) for the progenitors with masses of $M_{\rm ZAMS}=138 \msun$. Then, for progenitor stars with $138<M_{\rm ZAMS} \leq 153 \msun$, the final He core mass decreases a bit below the PSN limit and the star experience a PPSN instead. The limit for PSN is exceeded again for progenitors with initial masses $M_{\rm ZAMS}>153 \msun$. In the revised PSN variant, the limit for PSN is exceeded for most massive progenitors with $M_{\rm ZAMS}>250 \msun$. As in our binary evolution cosmological simulations (Section \ref{sec: Binary_Evol}), we limit the initial stellar mass to $M_{\rm ZAMS}=200 \msun$.  Therefore, a PSN in the revised treatment does not play a role in those results.
Stars with 10\%$Z_{\odot}$ do not experience PSN as the stellar winds reduce the He core mass below the PSN limit. However, in the strong PPSN/PSN model most massive stars may be a subject of PPSN. Due to the overlap of the stellar winds with mass reduction in PPSN the remnant masses of 10\%$Z_{\odot}$ stars is constantly $\sim 45 \msun$ for all the progenitors with initial masses $M_{\rm ZAMS}>105\msun$. On the other hand, the results for revised PSN model do not differ much, as mass loss with stellar winds allows for the formation of BH only to $\sim 47\msun$ in the wide range of $105 \msun < M_{\rm ZAMS} < 250 \msun$. For 100\%$Z_{\odot}$ stars do not experience PPSN nor PSN in both PSN variants as the strong stellar winds allows for the formation of BH with maximum mass of $\sim 20\msun$.

\subsection{Lower mass gap} \label{sec:MassGap}

\begin{figure*}
	\includegraphics[width=12 cm]{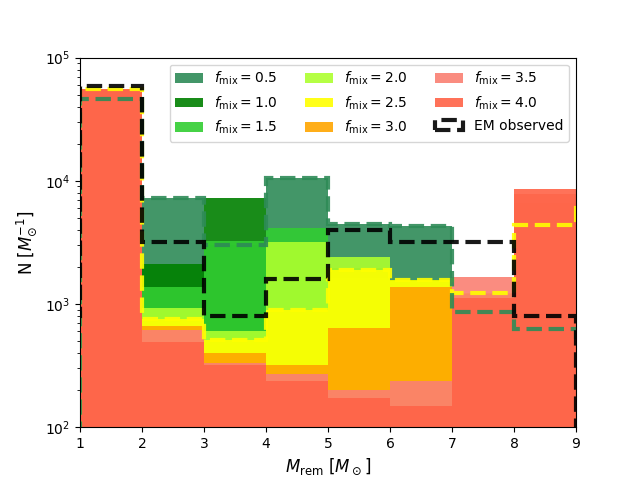}
    \caption{Histogram (binsize $1\msun$) of the remnant masses of the probe of $10^5$ single stars for adopted initial mass function for eight new SN model prescriptions with $f_{\rm mix}$ in a range 0.5-4.0 and $Z=1.0 Z_{\odot}$. Results for $M_{\rm crit}=5.75M_\odot$ adopted for this study (see Eq. \ref{eq:mrem}).}
    \label{fig:MassGap}
\end{figure*}

Figure \ref{fig:MassGap} demonstrates how the lower mass gap changes with the assumed value of $f_{\rm mix}$ parameter corresponding to different convection timescales growth (from rapid to delayed). These results are for high metallicity $100\% Z_{\odot}$. In this paper we assume $f_{\rm mix}$ takes a value from the range: 0.5 to 4.0 as suggested by \cite{Fryer2022} (see Sec. \ref{sec: Method}. In the Figure we show results for eight $f_{\rm mix}$ values chosen from this range. The presented histogram, with a bin size of $1\msun$, is made from the study of $10^5$ single stars (for each SN model) with their initial masses $5\msun < M_{\rm ZAMS} <150 \msun$ generated from three broken initial mass function \cite{1993MNRAS.262..545K,2002Sci...295...82K} with power-law exponent for massive stars (M$_{\rm ZAMS}>1\msun$) equal $\alpha_3=-2.3$. The distribution is shown in the range limited to remnant masses $1<M_{\rm rem}<9 \msun$ in order demonstrate clearly the lower mass gap region.

In the same Figure \ref{fig:MassGap}, we plot (black dashed line) the distribution NSs and low mass BHs ($<9\msun$) observed in electromagnetic spectrum using database collected by LVK \href{https://ligo.northwestern.edu/media/mass-plot/index.html?fbclid=IwAR3NOfvIWxzFtt-GyiSQThp2jOv2HW-yYmT-hbRfmonsahbn1V_gzAswPII}{LIGO-Virgo Mass Plot} with provided references on system parameters: \cite{Deller2012,Heida2017,Alsing2018,Thompson2019,Gieses2019,Ferdman2020,Fonseca2021,Jayasinghe2021,Haniewicz2021}.

For all tested models, we observe a peak of NSs with masses in the range $M_{\rm rem} \in 1-2 \msun$. The distribution for each $f_{\rm mix}$ varies significantly. In the model with the lowest $f_{\rm mix}=0.5$ (the most delayed convection growth), we get a high fraction ($\sim 10^4 \msun^{-1}$) of compact object remnants, massive NSs and low mass BHs, within the range of $2\msun \leq M_{\rm rem}<7 \msun$. The corresponding number of more massive compact objects decreases by around an order of magnitude. The distribution of remnant masses for the most rapid model, $f_{\rm mix}=4.0$ behaves in the opposite way. After the NSs peak, there is a deep drop by $\sim 2$ orders of magnitude for compact objects in the mass bin $2\msun \leq M_{\rm rem}<7 \msun$. The number of BHs increases by 1-2 orders of magnitude for $M_{\rm rem} \in 7-9 \msun$ compared to mass range $M_{\rm rem} \in 6-7 \msun$. The increase in the fraction of objects within two most massive plotted bins, as already mentioned in Sec. \ref{sec:ZAMS}, is due to peak in the convection of the rapid convection timescale for $f_{\rm mix}=4.0$ occuring earlier when the ram pressure of the infalling star is higher. The pre-SN star goes through direct collapse to a BH, with only mass loss in neutrino flux, leaving more massive remnants than in case of successful SN explosion (e.g. $f_{\rm mix}=0.5$).  The results for other $f_{\rm mix}$ alter the width and depth of the gap region.  The models with $f_{\rm mix}$ values in the middle of possible range, e.g. $f_{\rm max}$=2.5, are promising as they reconstruct the reduced number of observed compact objects withing the range $2-5 \msun$ but also do not completely prevent their formation. In this respect, such models are most compatible with small database of NSs and low mass BHs observed in electromagnetic spectrum (black dashed line). Note that here we compare results from simulations with the observed population of compact objects without taking into account many possible biases. 

For comparison, apart from the results for $M_{\rm crit}=5.75M_\odot$ adopted for this study (see Eq. \ref{eq:mrem}), in Figure \ref{fig:MassGap_mcrit4.75} we provide same results but for different critical mass of carbon oxygen core for a BH formation, $M_{\rm crit}=4.75M_\odot$. Different choice of $M_{\rm crit}$ parameter value may affect the range of produced mass gap. 

\section{Binary evolution} \label{sec: Binary_Evol}

In this section we present the impact of the new remnant mass prescriptions for cosmological population of DCO mergers formed via isolated binary evolution. Besides testing new formulas for remnant masses we also show results for two different RLOF stability criteria{ and two PSN treatment, see method description in Sec. \ref{sec: StarTrack} for more details. }In Sections \ref{subsec: mass_distribution} and \ref{subsec:mass_ratio}, we present mass and mass ratio distributions of DCO mergers respectively. In Section \ref{subsec: local_densities}, we provide estimates of local merger rate densities for different types of DCO mergers.

\subsection{Mass distribution of DCO mergers} \label{subsec: mass_distribution}

\begin{figure*}
	\includegraphics[width=8.6 cm]{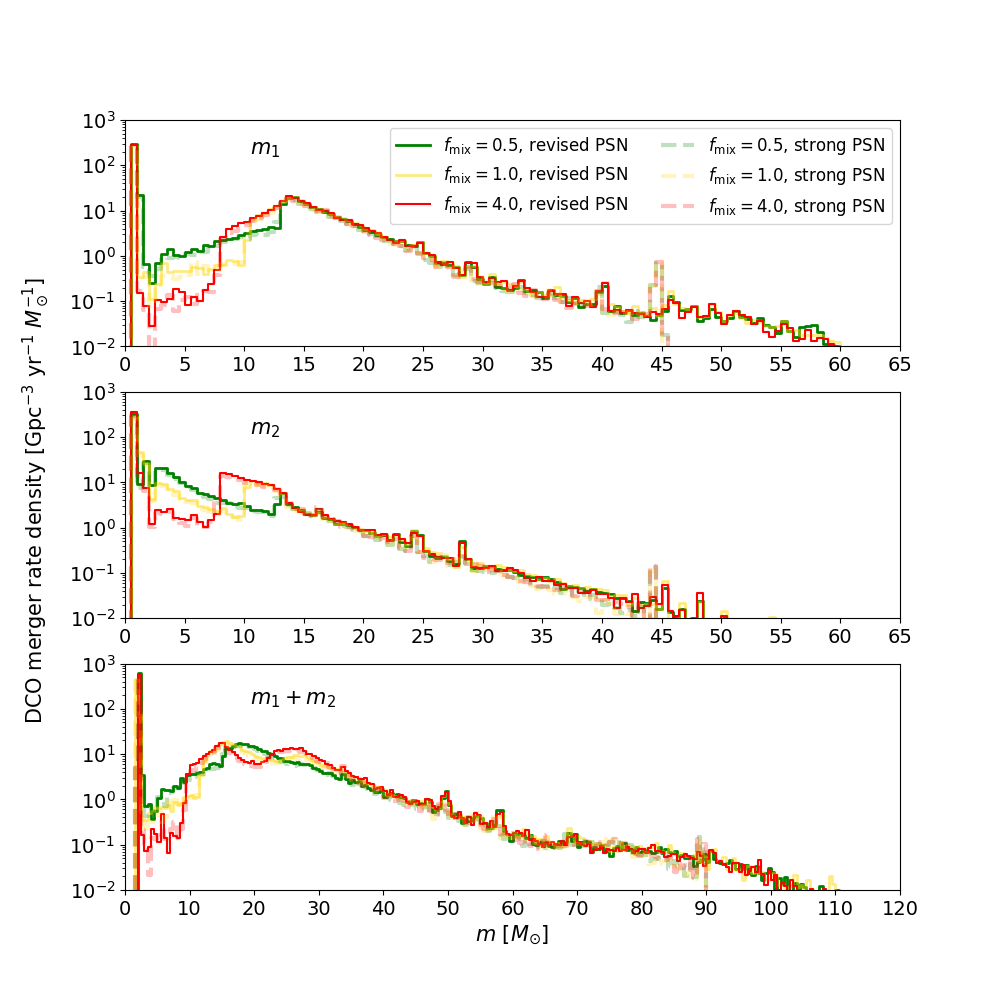}
	\includegraphics[width=8.6 cm]{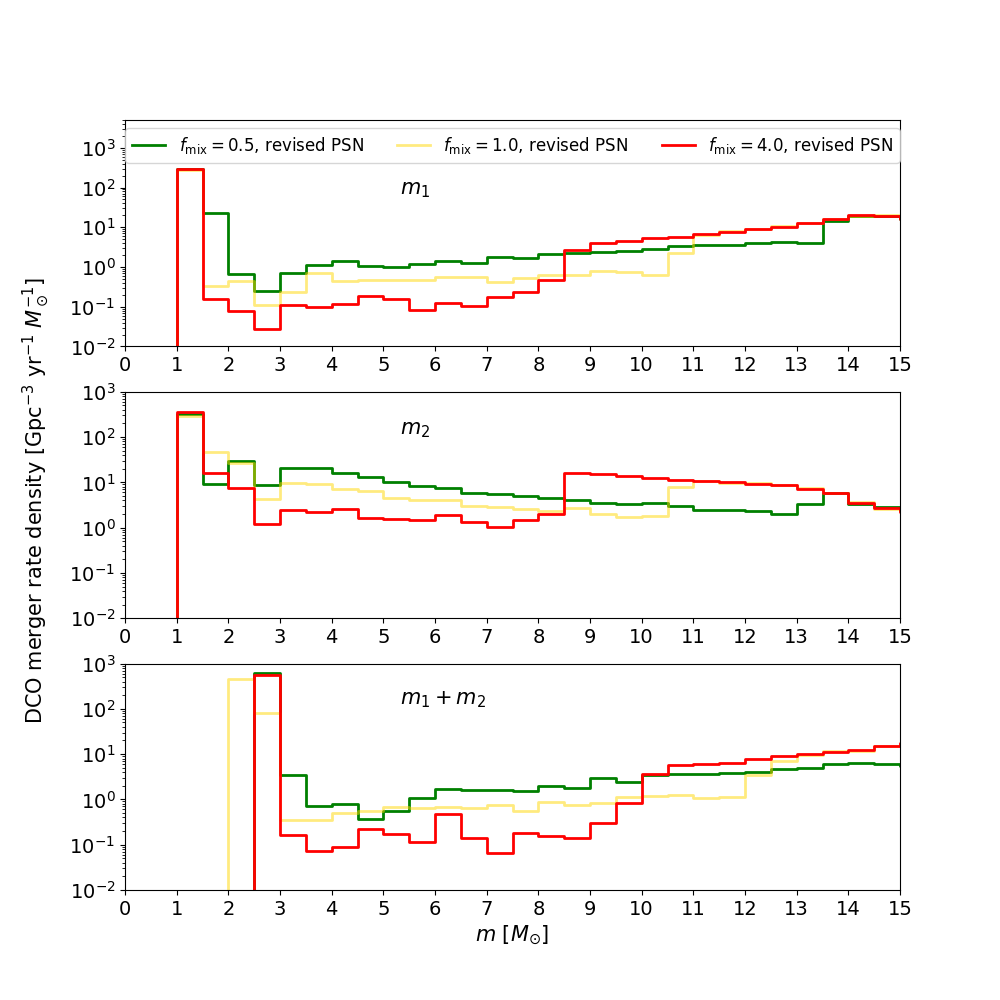}
    \caption{Intrinsic mass distribution of all types (NS-NS, BH-NS, BH-BH) mergers at low redshifts ($z<1.0$) for standard common envelope development criteria (BH-BH mergers forms via common envelope evolution). Results for three new remnant mass formulas with f$_{\rm mix}=0.5;1.0$ and 4.0 ($m_1 \geq m_2$). On the left the whole spectrum of masses for two PSN models: revised PSN (solid lines) and strong PPSN/PSN (dashed, semitransparent lines). On the right mass range limited to $15 \msun$ (only revised PSN model). Note that our models allow for various depths of the lower mass gap, from deep (red line, the most rapid SN) to shallow (green line, the most delayed SN) gap depending on the development timescale of neutrino supported convection SN engine. Full description in Section \ref{sec: Binary_Evol}. }
    \label{fig:MassDistribution_DCO_CE}
\end{figure*}

\begin{figure*}
    \includegraphics[width=8.6 cm]{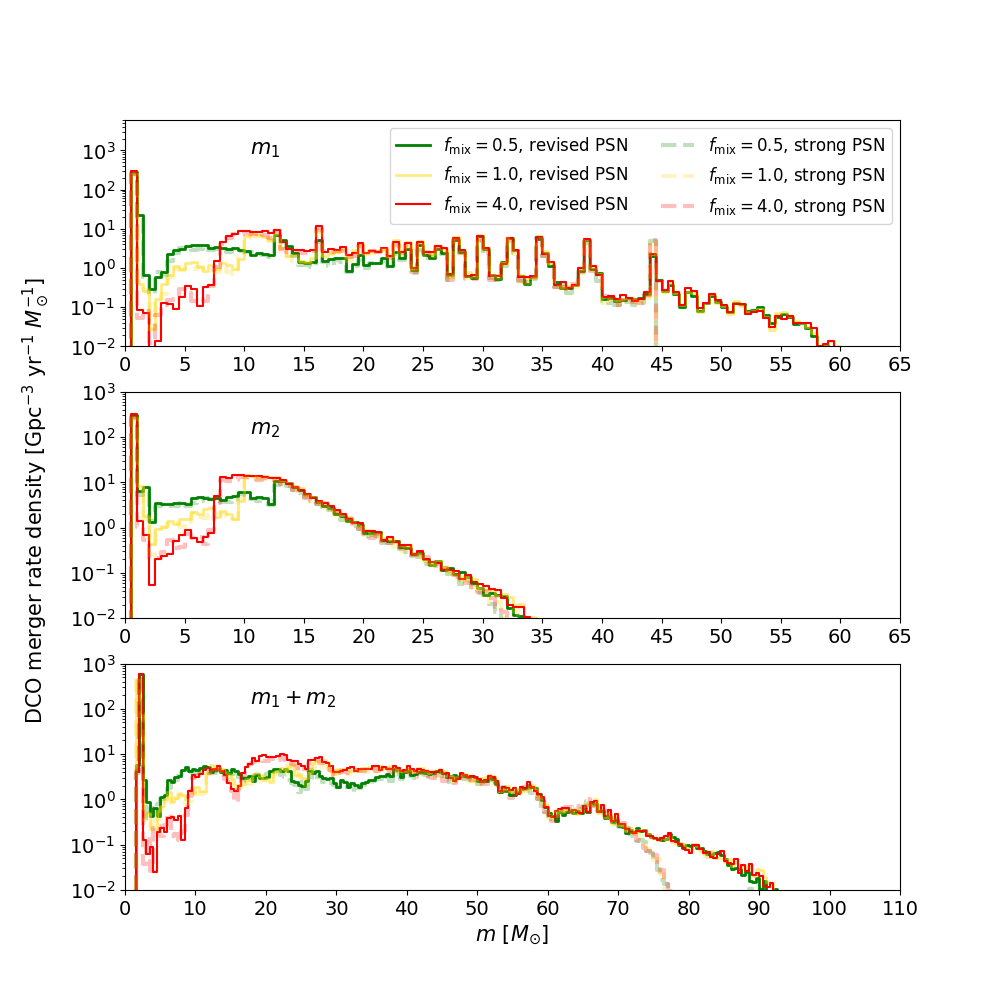}
    \includegraphics[width=8.6 cm]{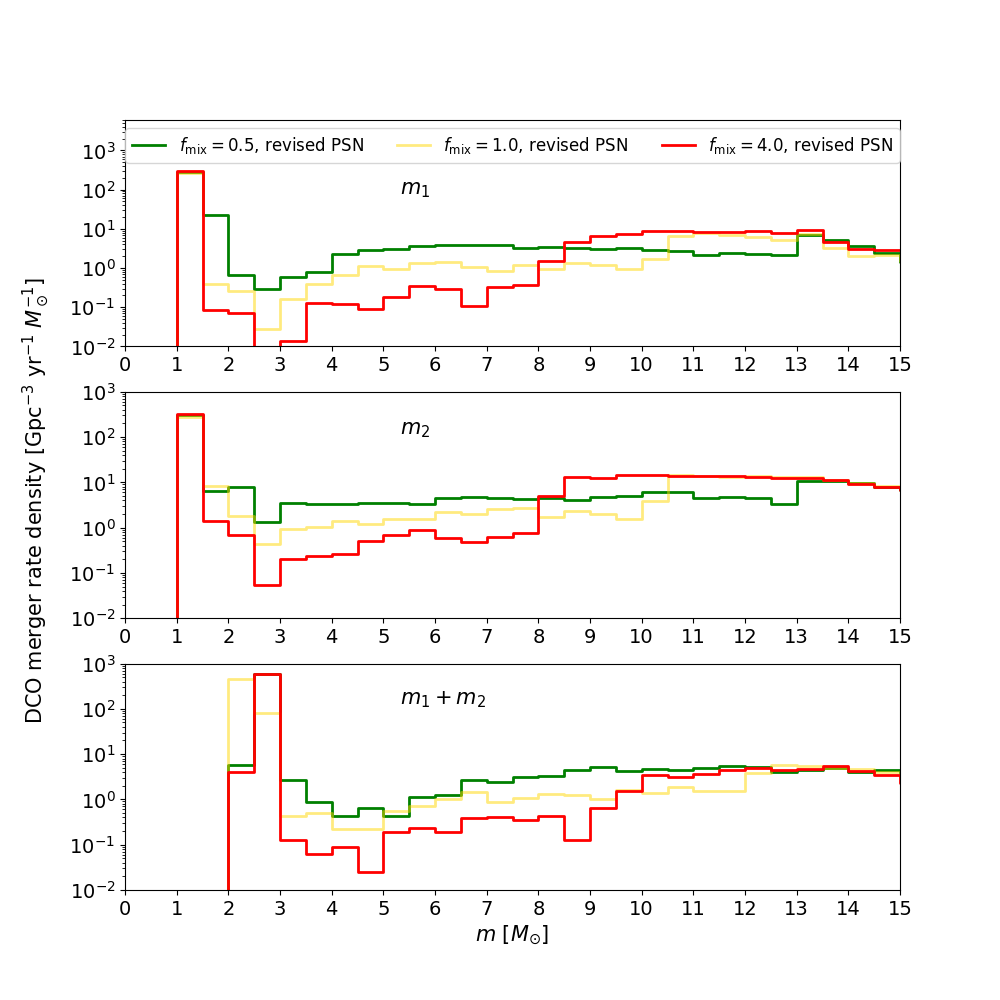}
    \caption{Intrinsic mass distribution of all types (NS-NS, BH-NS, BH-BH) mergers at low redshifts ($z<1.0$) for the revised criteria for mass transfer stability (BH-BH mergers forms via stable RLOF). Results for three new remnant mass formulas with f$_{\rm mix}=0.5;1.0$ and 4.0 ($m_1 \geq m_2$). On the left the whole spectrum of masses for two PSN models: revised PSN (solid lines) and strong PPSN/PSN (dashed, semitransparent lines). On the right mass range limited to $15 \msun$ (only revised PSN model). Note that our models allow for various depths of the lower mass gap, from deep (red line, the most rapid SN) to shallow (green line, the most delayed SN) gap depending on the development timescale of neutrino supported convection SN engine. Full description in Section \ref{sec: Binary_Evol}. }
    \label{fig:MassDistribution_DCO_Pav}
\end{figure*}

In Figures \ref{fig:MassDistribution_DCO_CE} and \ref{fig:MassDistribution_DCO_Pav}, we present combined intrinsic distributions of remnant masses for all types of DCO which merge at redshift $z<1.0$. The plots in Figure \ref{fig:MassDistribution_DCO_CE} are results for the standard CE {\tt StarTrack} development criteria, in which a vast majority of BH-BH binaries form via CE evolution. The plots in Figure \ref{fig:MassDistribution_DCO_Pav} show results for the revised CE development criteria, in which a majority of BH-BH binaries form via stable RLOF channel without any CE phase. For both figures, we provide plots with the entire spectrum of masses and with a mass range limited to low masses, m$<15\msun$ for a better visibility of the lower mass gap region. We show distributions of the more massive merger component (the primary, $m_1$), less massive merger component (secondary, $m_2$) and a sum of the two masses ($m_1+m_2$). In every plot we provide results for three cases of new SN models: two extreme cases for convection growth timescales, the most delayed with $f_{\rm mix}=0.5$, the most rapid with $f_{\rm mix}=4.0$ and intermediate case for $f_{\rm mix}=1.0$. We also plot results for the two adopted limits for PSN. 

\begin{figure*}
	\includegraphics[width=12.5 cm]{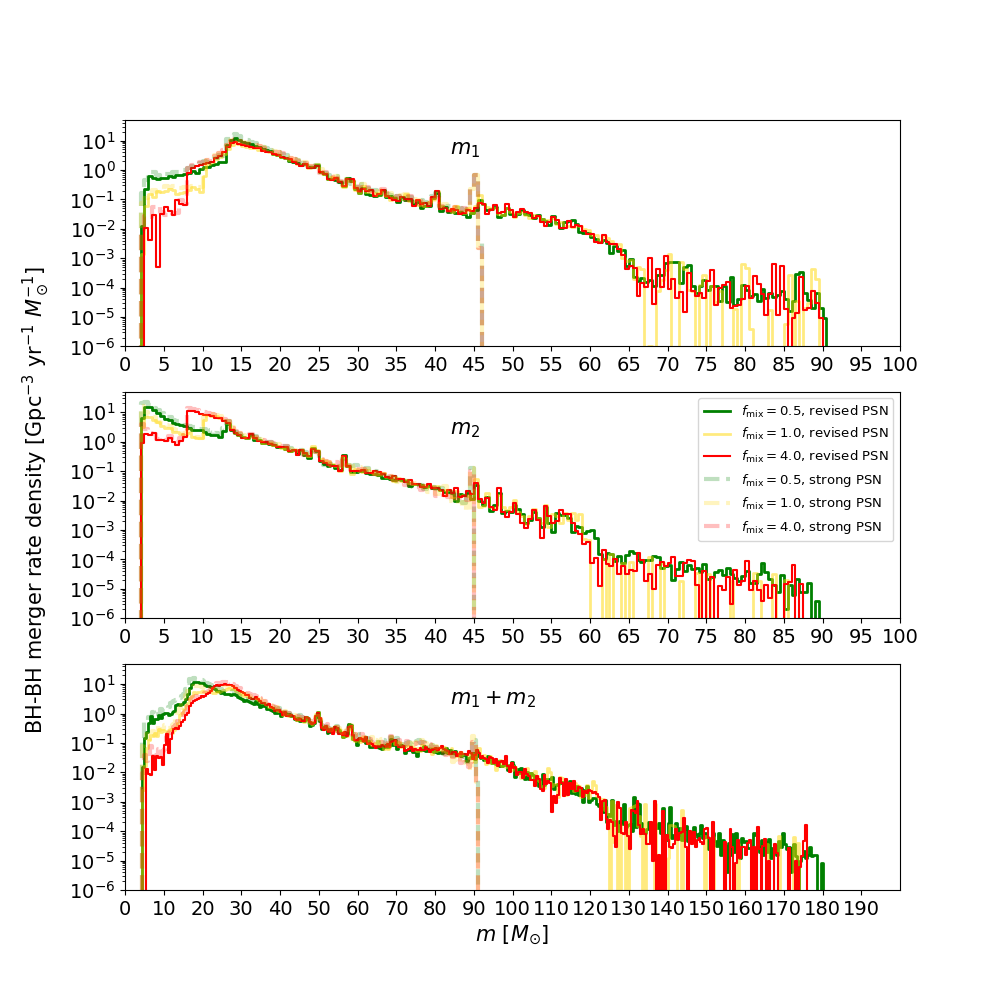}
	    \caption{Intrinsic mass distribution of BH-BH mergers at low redshifts ($z<1.0$) for standard common envelope development criteria (BH-BH mergers forms via common envelope evolution) divided into bins with $0.5 \msun$ size. Results for three new remnant mass formulas with f$_{\rm mix}=0.5;1.0$ and 4.0 ($m_1 \geq m_2$) for two PSN models: revised PSN (solid lines) and strong PPSN/PSN (dashed, semitransparent lines). Note that our models allow for various depths of the lower mass gap, from deep (red line, the most rapid SN) to shallow (green line, the most delayed SN) gap depending on the development timescale of neutrino supported convection SN engine. Full description in Section \ref{sec: Binary_Evol}. }
    \label{fig:MassDistribution_BHBH_CE}
\end{figure*}

\begin{figure*}
    \includegraphics[width=12.5 cm]{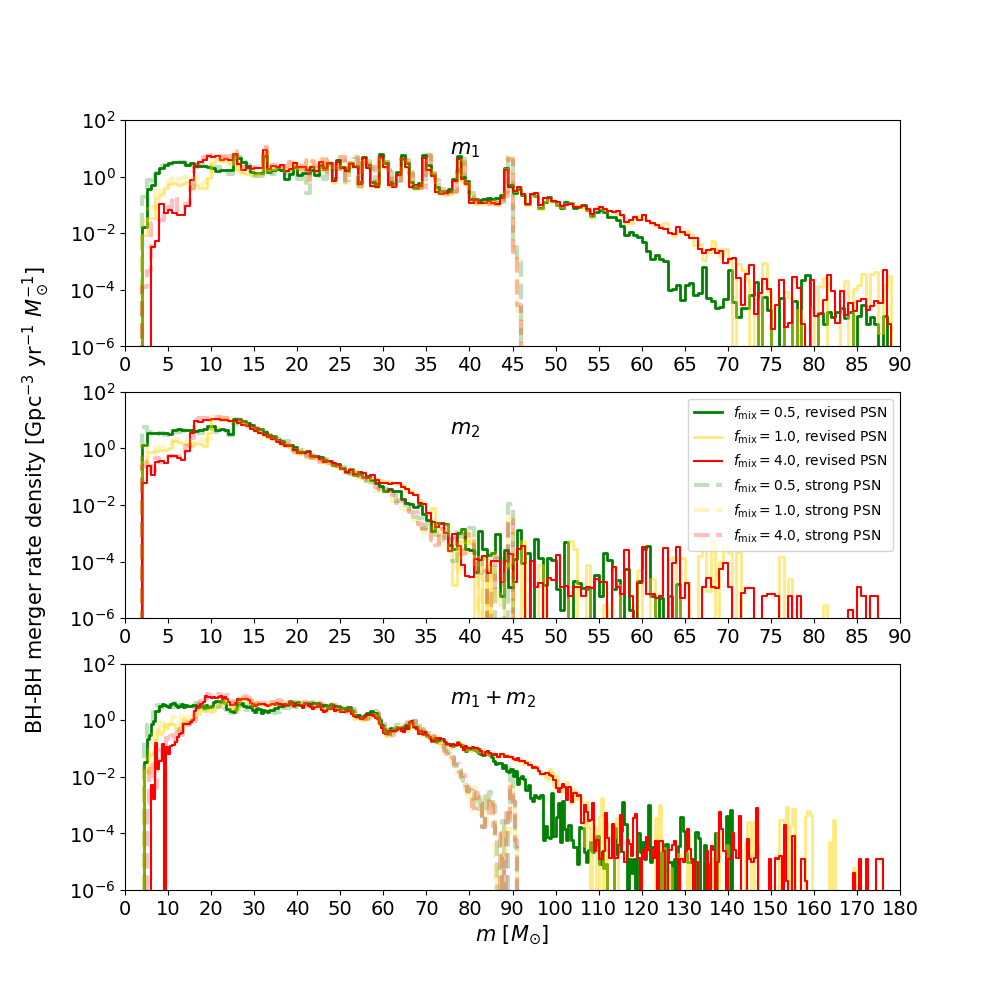}
    \caption{Intrinsic mass distribution of BH-BH mergers at low redshifts ($z<1.0$) for revised mass transfer stability criteria (BH-BH mergers forms via stable RLOF) divided into bins with $0.5 \msun$ size. Results for three new remnant mass formulas with f$_{\rm mix}=0.5;1.0$ and 4.0 ($m_1 \geq m_2$) for two PSN models: revised PSN (solid lines) and strong PPSN/PSN (dashed, semitransparent lines). Note that our models allow for various depths of the lower mass gap, from deep (red line, the most rapid SN) to shallow (green line, the most delayed SN) gap depending on the development timescale of neutrino supported convection SN engine. On the right mass range limited to $15 \msun$. Full description in Section \ref{sec: Binary_Evol}.}
    \label{fig:MassDistribution_BHBH_Pav}
\end{figure*}

In all distributions (for all remnant mass models and both CE development criteria), there is a large peak in the primary mass $m_1$ distribution in the mass range corresponding to typical masses of NSs. In the primary mass distribution for the model with f$_{\rm mix}=0.5$, there is significant fraction of more massive NSs ($m_1 \in 1.5-2.0 \msun$), which we do not observe for two other tested SN models (corresponding to more rapid convection timescale growth). For the model with f$_{\rm mix}=0.5$, the fraction of DCO mergers with their total masses in range: $m_1+m_2 \in 3-4\msun$ (massive NS-NS mergers) is about one order of magnitude higher than in other models. This may be important feature for studying the origin of systems such as GW190425 \citep{massive_NSNS2020} classified as a massive NS-NS merger with its component masses: $m_1 \in 1.60 - 1.87\msun$ and $m_2 \in 1.46 - 1.69 \msun$ for low-spin prior (or $m_1 \in 1.61 - 2.52$ and $m_2 \in 1.12-1.68 \msun$ with high-spin prior assumption).

The fraction of DCO mergers with components within the lower mass gap and slightly beyond it is systematically about one order of magnitude lower for the most rapid tested model with $f_{\rm mix}=4.0$ than for most delayed one $f_{\rm mix}=0.5$. The results corresponding to intermediate case of convection growth timescale ($f_{\rm mix}=1.0$) are usually somewhere between the two extreme cases. After exceeding the masses of $m_1,m_2 \geq 8\msun$ there is a reversal of trends in the behavior of extreme SN models distributions. The  fraction of DCO with $m_{1},m_{2} \in 9-15 \msun$ for $f_{\rm mix}=4.0$ is a few times larger than in $f_{\rm mix}=0.5$. In the case of $f_{\rm mix}=1.0$, the fraction of DCO mergers remains low for a wide range of masses $m_{1},m_{2} \in 2-10 \msun$ and increases by about an order of magnitude for $m_{1},m_{2} \gtrsim 11 \msun$. For the reasons already mentioned in Section \ref{sec: Single_Evol}, more rapid SN models (e.g. f$_{\rm mix}=4.0$) with successful explosions are unlikely to produce remnants with masses in the lower mass gap. Instead stars either explode rapidly to form a NS below the gap or collapse directly to a BH with a mass above the gap.  The influence of the adopted SN model is visible in the region of $m_1,m_2 \lesssim 17 \msun$ ($m_1+m_2 \lesssim 35 \msun$). After that threshold, massive stars in all tested models end their lives in a direct collapse to a BH, without a successful SN explosion. 
Different adopted approaches for the PSN limit doesn't affect much the general picture of DCO mass distribution. The main difference is a tail of BHs with masses larger than $45 \msun$ in the model with revised PSN limit while for strong PPSN/PSN model there is a narrow peak around $45 \msun$ due to mass reduction by PPSN. 
The difference in the distribution of DCO mergers between two CE development criteria (Fig. \ref{fig:MassDistribution_DCO_CE} and \ref{fig:MassDistribution_DCO_Pav}) is especially visible for massive systems ($m_{1},m_{2} \gtrsim 20 \msun$). BH-BH mergers formed in stable RLOF scenario (Fig. \ref{fig:MassDistribution_DCO_Pav}) are characterised by significantly larger average total mass of DCO mergers, what was already noticed and explained by other recent studies, for example \cite{vanSon2021} and \cite{Belczynski2022b}. Massive BH-BH mergers are more likely to form via stable RLOF evolution as their progenitors avoid stellar merger while entering CE with Hertzsprung gap star donor \citep{Olejak2021a}. Different approaches to the mass and orbital angular momentum loss mechanisms during stable and unstable RLOF also strongly affect mass ratio of BH-BH mergers (See Sec. \ref{subsec:mass_ratio}).
Additionally, in Figures \ref{fig:MassDistribution_BHBH_CE} and \ref{fig:MassDistribution_BHBH_Pav}, we show  mass distribution only for BH-BH mergers for two RLOF stability criteria. BH-BH mergers strongly dominate the detected GW signals thusfar and therefore they are sometimes analyzed separately from other types of DCO mergers.

\subsection{Mass ratio distribution of BH-BH and BH-NS mergers} \label{subsec:mass_ratio}

\begin{figure*}
	\includegraphics[width=8.5 cm]{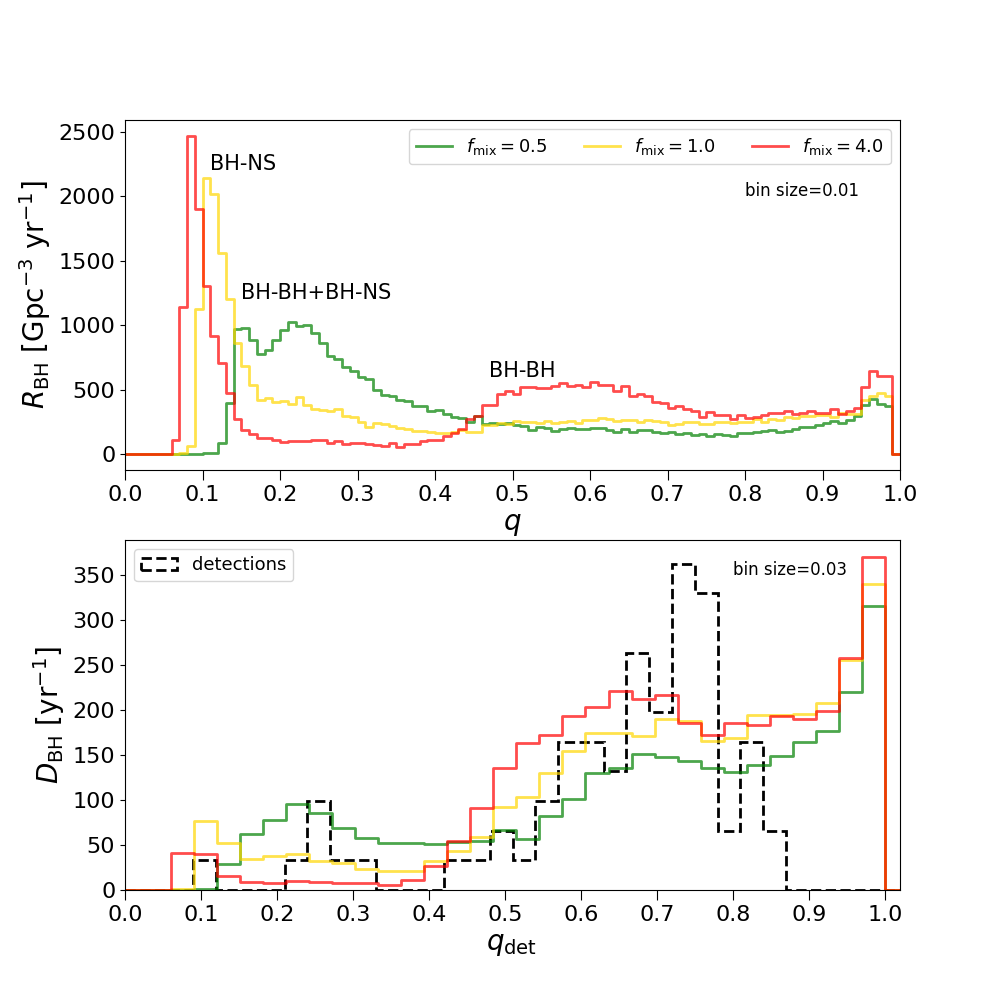}
	\includegraphics[width=8.5 cm]{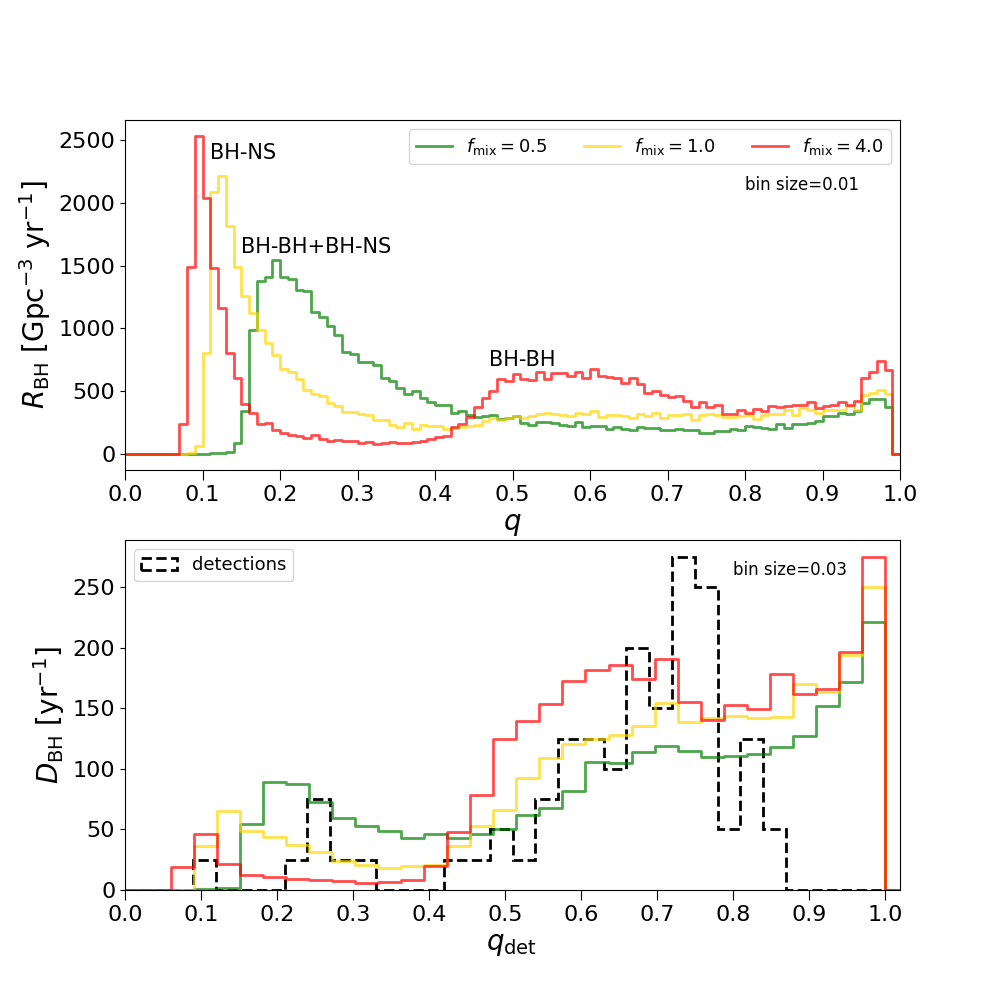}
    \caption{Mass ratio distribution for BH-BH and BH-NS mergers at low redshifts ($z<1.0$) for the standard common envelope development criteria (BH-BH mergers formed via common envelope). On the upper panel intrinsic results (not redshifted nor detection-weighted). On the bottom panel detection-weighted results together with the parameters of detected population of BH-BH and BH-NS mergers (O1+O2+O3 runs). \textbf{On the left:} distribution for revised PSN model; \textbf{on the right:} distribution for strong PPSN/PSN.}
    \label{fig:MassRatioCE}
	\includegraphics[width=8.5 cm]{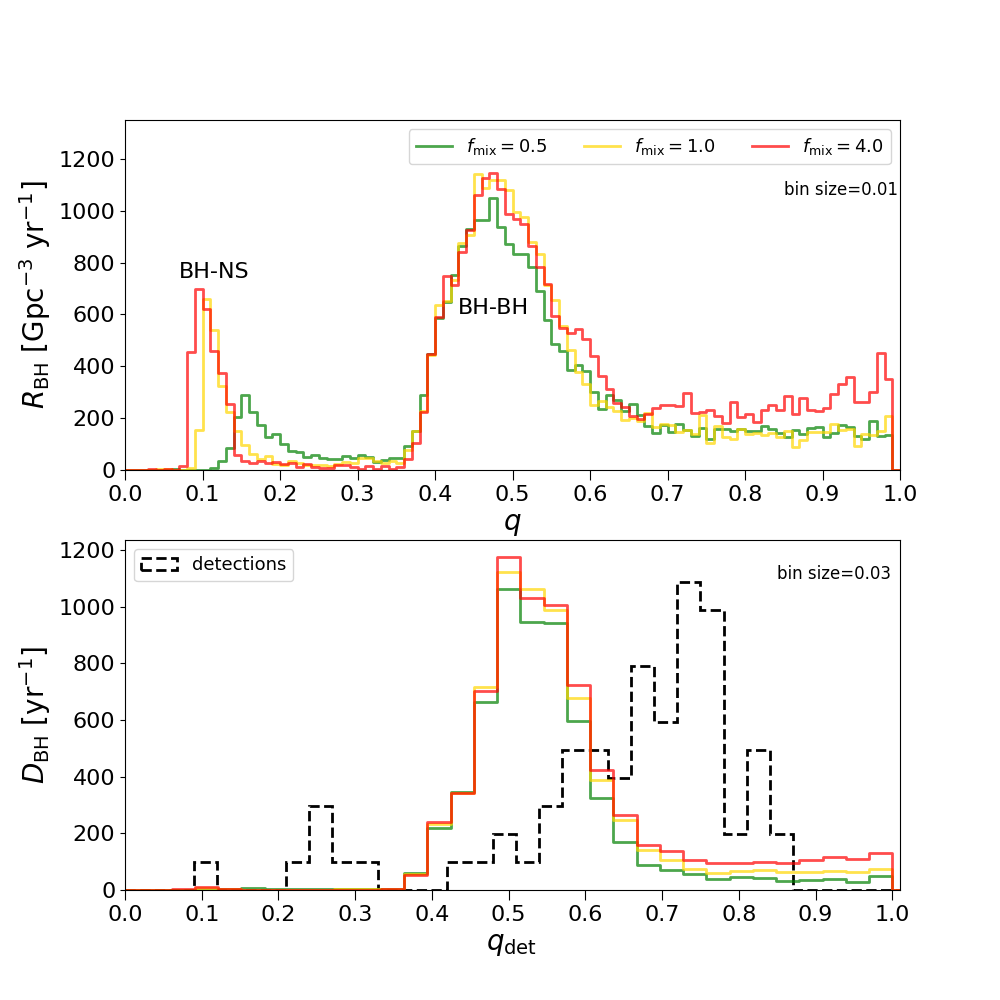}
	\includegraphics[width=8.5 cm]{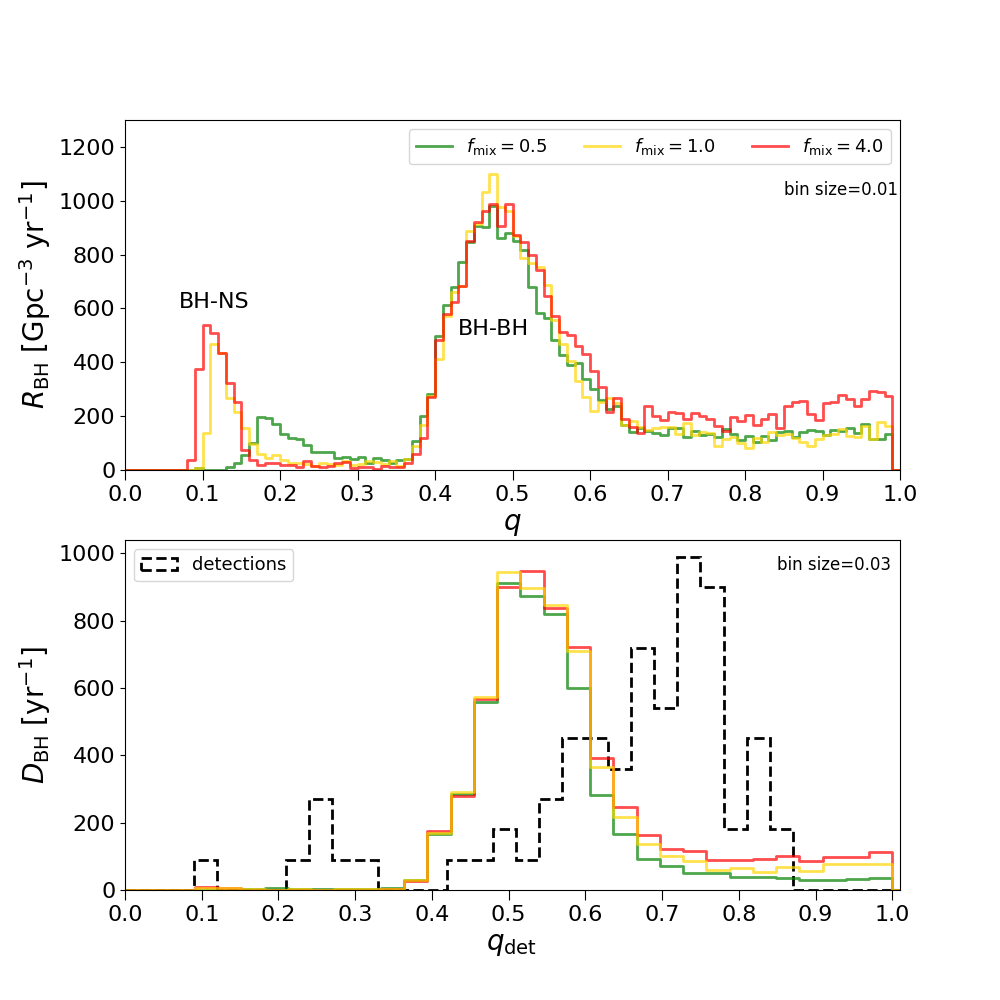}
    \caption{Mass ratio distribution for BH-BH and BH-NS mergers at low redshifts ($z<1.0$) for the revised mass transfer stability criteria (BH-BH mergers formed via stable RLOF). On the upper panel intrinsic results (not redshifted nor detection-weighted). On the bottom panel detection-weighted results together with parameters of detected population of BH-BHs and BH-NS mergers (O1+O2+O3 runs). \textbf{On the left:} distribution for revised PSN; \textbf{on the right:} distribution for strong PPSN/PSN.}
    \label{fig:MassRatioPav}
\end{figure*}

In the Figures \ref{fig:MassRatioCE} and \ref{fig:MassRatioPav}, we present the mass ratio  distributions of BH-BH and BH-NS mergers (combined) obtained for three examples of new SN models: $f_{\rm mix}=0.5$, $f_{\rm mix}=1.0$ and $f_{\rm mix}=4.0$ similarly as in \ref{subsec: mass_distribution}.
The mass ratio $q$ is defined here as the mass of the secondary (less massive) to mass of the primary (more massive) component of merger ($q = \frac{m_2}{m_1}$).
In Figure \ref{fig:MassRatioCE}, we plot the results for the standard CE development criteria, and in Figure \ref{fig:MassRatioPav}, the results for the revised criteria under which BH-BH mergers form mostly through stable RLOF. For both, the standard and revised criteria, we show two separate figures corresponding to two different PSN variants (see Sec. \ref{sec: StarTrack}). On the left we show results for the revised PSN model and, on the right, results for the strong PPSN/PSN model. On the top panel of the figures, we present the intrinsic (not redshifted nor detection-weighted) mass ratio distribution of BH-BH and BH-NS mergers population (combined) at redshift $z<1.0$. On the bottom panel we also plot combined mass ratio distribution of BH-BH and BH-NS but weighted by detection biases according to a method described in Section \ref{subsec: Detections} (only signals with estimated SNR$>8$). The detection-weighted results are plotted together with a distribution built of publicly announced during O1+O2+O3 runs parameters of BH-BH and BH-NS mergers with a black dashed line \citep{2016PhRvX...6d1015A,LIGO2019a,Abbott_2019,LIGO2019b,Abbott_2021,LIGOfullO3population2021}.\\ 

In Table \ref{tab:mass_ratio_fractions}, for all tested physical variants (see Sec. \ref{sec: Method}), we provide total fractions of unequal mass mergers with their mass ratios: $q<0.3$ and $q<0.5$.
The first column is a row number, in the second column we specify the adopted criteria for CE development: standard or revised. In the third column we define used PSN model: strong (which limits BH mass to $\sim 45 \msun$) or revised (which allows for more massive BHs formation). In the fourth column (SN model) we specify adopted mixing parameter value for the new remnant mass formula: $f_{\rm mix}=0.5;1.0$ or 4.0. In last four columns we give total fractions of unequal mass ratio mergers, first for intrinsic distributions $q_{\rm intr}$ (column fifth and sixth) and next for the detection weighted results $q_{\rm det}$ (column seventh and eighth).

\begin{table*} 
     $$ 
         \begin{array}{lllccccc}
            \hline
            \noalign{\smallskip}
            \mbox{No.} & \mbox{CE criteria} & \mbox{PSN limit}  & \mbox{SN model}  & { q_{\rm intr}<0.5}  & { q_{\rm intr}<0.3} & { q_{\rm det}<0.5}  & { q_{\rm det}<0.3}  \\
            \noalign{\smallskip}
            \hline
            \noalign{\smallskip}
            \mbox{1.}& \mbox{Standard} & 90\msun & f_{\rm mix}=0.5 & 0.68 & 0.44 & 0.24 & 0.13\\
            \mbox{2.}& \mbox{Standard} & 90\msun & f_{\rm mix}=1.0 & 0.61 & 0.47 & 0.15 & 0.08\\
            \mbox{3.}& \mbox{Standard} & 90\msun & f_{\rm mix}=4.0 & 0.43 & 0.32 & 0.10 & 0.04\\
            \noalign{\smallskip}
            \hline
            \noalign{\smallskip}
            \mbox{4.}& \mbox{Standard} & 45\msun & f_{\rm mix}=0.5 & 0.68 & 0.43 & 0.26 & 0.13\\
            \mbox{5.}& \mbox{Standard} & 45\msun & f_{\rm mix}=1.0 & 0.62 & 0.48 & 0.17 & 0.10 \\
            \mbox{6.}& \mbox{Standard} & 45\msun & f_{\rm mix}=4.0 & 0.43 & 0.32 & 0.11 & 0.04\\
            \noalign{\smallskip}
            \hline
            \noalign{\smallskip}
            \mbox{7.}& \mbox{Revised} & 90\msun & f_{\rm mix}=0.5 & 0.45 & 0.08 & 0.28 & 0.003 \\
            \mbox{8.}& \mbox{Revised} & 90\msun & f_{\rm mix}=1.0 & 0.66 & 0.10 & 0.26 & 0.003 \\
            \mbox{9.}& \mbox{Revised} & 90\msun & f_{\rm mix}=4.0 & 0.57 & 0.11 & 0.24 & 0.003 \\
            \noalign{\smallskip}
            \hline
            \noalign{\smallskip}

            \mbox{10.}& \mbox{Revised} & 45\msun & f_{\rm mix}=0.5 & 0.44 & 0.07 & 0.26 & 0.003 \\
            \mbox{11.}& \mbox{Revised} & 45\msun & f_{\rm mix}=1.0 & 0.63 & 0.10 & 0.23 & 0.003 \\
            \mbox{12.}& \mbox{Revised} & 45\msun & f_{\rm mix}=4.0 & 0.54 & 0.11 & 0.21 & 0.003 \\
            \hline
         \end{array}
     $$ 
     \caption{
     Fractions of unequal mass BH-BH and BH-NS mergers (combined) in different tested models. The columns: CE criteria stands for adopted standard or revised CE development treatment; PSN limit stands for adopted strong ($\sim 45 \msun$) or revised ($\sim 90 \msun$) PSN limit; SN model stands for different adopted $f_{\rm mix}$ parameters (convection growth time); $q_{\rm int}<X$ is the intrinsic fraction of mergers with mass ratio less than $X$ while $q_{\rm det}<Y$ is the fraction of mergers weighted by detection with mass ratio less than $Y$.}    
     \label{tab:mass_ratio_fractions}
\end{table*}

\subsubsection{CE evolution}

The intrinsic mass ratio distribution of BH-BH and BH-NS mergers (combined) in the case of CE evolution channel (top panel of Fig. \ref{fig:MassRatioCE}) is significantly affected by the adopted model for the SN explosion. In the distribution for our model with $f_{\rm mix}=4.0$, which produces deep and wide lower mass gap, there are three peaks: a very high, thin peak at $q \approx 0.15$ composed of BH-NS mergers, a wide peak at $q \in 0.5-0.7$ and a slight peak at $q \approx 1.0$ both primarily composed of BH-BH mergers. The distribution for the model with $f_{\rm mix}=1.0$ is similar to that for the model with $f_{\rm mix}=4.0$ with a high peak for unequal mass BH-NS mergers.

In the mass ratio distribution for a model with $f_{\rm mix}=0.5$, there is a broad peak for unequal mass systems made of both BH-NS and BH-BH mergers with mass ratios $q \in 0.1-0.3$.  This SN model produces higher a fraction of massive NSs and low mass BHs within the lower mass gap via a successful SN explosion comparing to $f_{\rm mix}=4.0$ and $f_{\rm mix}=1.0$. Progenitors of BH-NS systems where both components are within the lower mass gap are expected to get high natal kicks at the time of a NS and BH formation. That makes such systems very easy to be disrupted or have orbits far too wide to merge in Hubble time. Therefore, despite producing massive NSs and low mass BHs, we still get mainly unequal mass BH-NS, even for a model with $f_{\rm mix}=0.5$. The models with $f_{\rm mix}=1.0$ and $f_{\rm mix}=4.0$ produce more BH-NS mergers due to a bigger fraction of massive BHs formed via direct collapse without getting significant kicks. In such systems a BH is usually few times more massive than a NS, what explains the origin of the high peak of very unequal mass mergers for the models with $f_{\rm mix}=1.0$ or $f_{\rm mix}=4.0$.  Assumption on PSN limit has negligible effect on mass ratio distribution.
 
\begin{figure*}
	\includegraphics[width=10 cm]{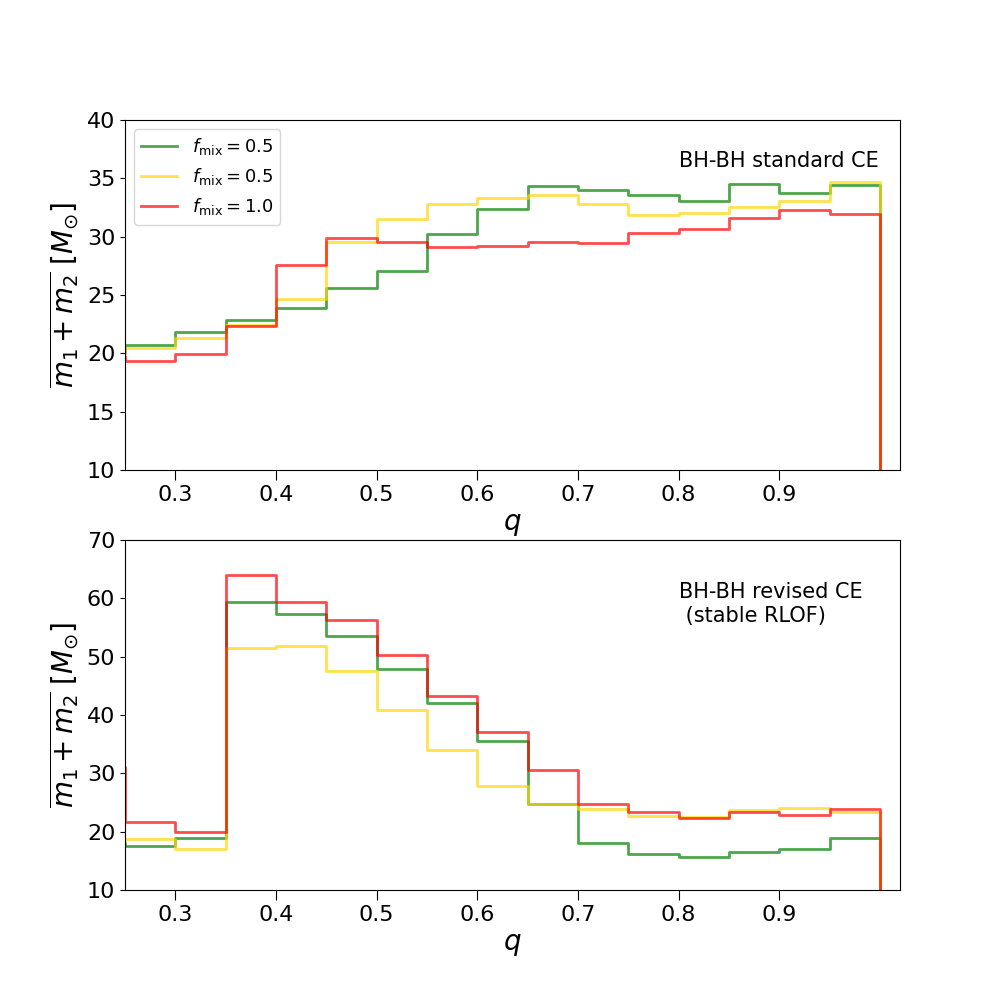}
    \caption{The relation between mass ratio $q$ and the average total mass $m_1+m_2$ of BH-BH mergers at $z<2.0$. On the top panel results for standard CE development criteria (formation mainly via CE), on the bottom panel results for revised CE development criteria (formation mainly via stable RLOF). Green line - model with $f_{\rm mix}=0.5$, yellow line - model with $f_{\rm mix}=1.0$ and red line - model with $f_{\rm mix}=4.0$. Results for revised PSN model.}
    \label{fig:q_vs_mtot}

	\includegraphics[width=10 cm]{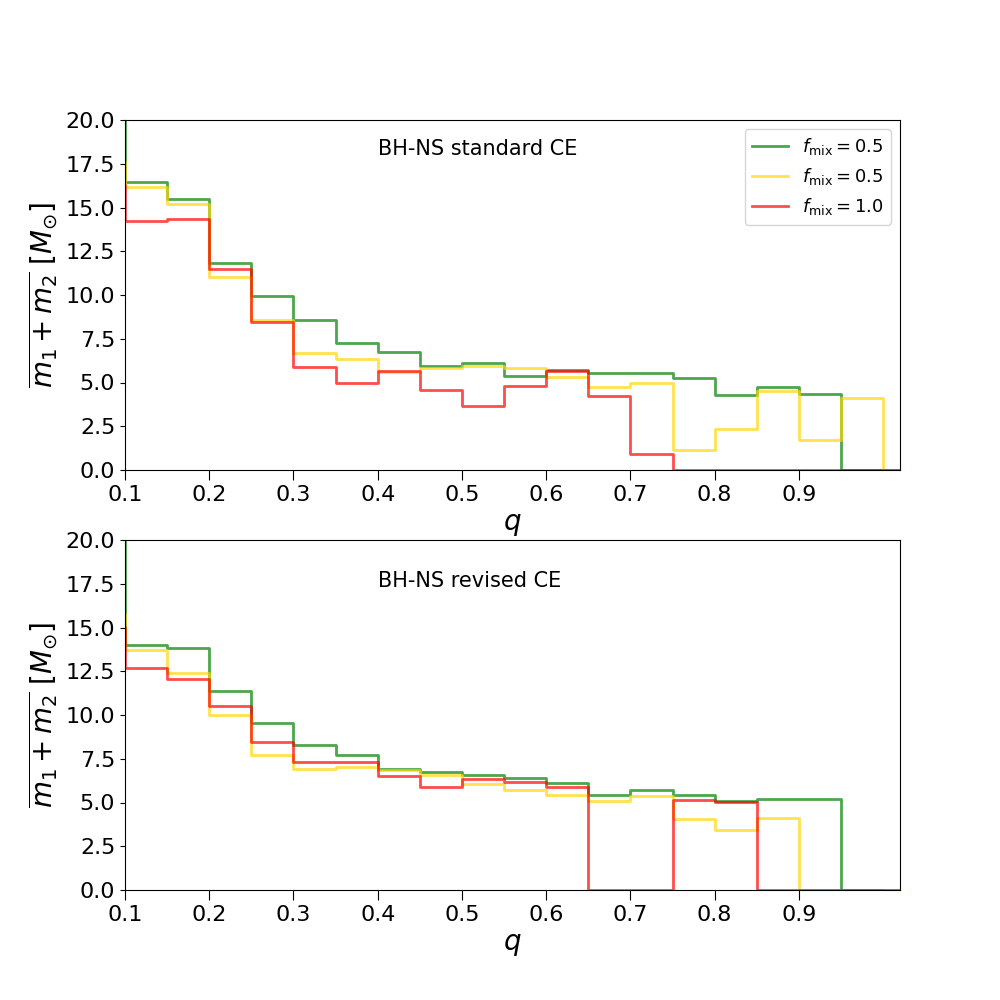}
    \caption{The relation between mass ratio $q$ and the average total mass $m_1+m_2$ of BH-NS mergers at $z<2.0$. On the top panel results for standard CE development criteria, on the bottom panel results for revised CE development criteria. Green line - model with $f_{\rm mix}=0.5$, yellow line - model with $f_{\rm mix}=1.0$ and red line - model with $f_{\rm mix}=4.0$. Results for revised PSN model.}
    \label{fig:q_vs_mtot_bhns}
\end{figure*}

Significant differences in intrinsic mass ratio distributions between SN models diminishes while weighting the result by detection biases. From the total BH-BH and BH-NS mergers population we choose only mergers with SNR$>8$. 
The observational distribution of detectable sources is governed by their SNR, which depends on the masses of compact objects in addition to other extrinsic parameters (like redshift, sky position, etc.).  The measured distribution of mass ratios from a population of detected binaries is significantly different from the intrinsic astrophysical distribution due to this selection bias. For stellar-mass compact objects, the SNR increases with the total mass of the binary for a fixed mass ratio. Also, for a given total mass, systems with equal mass components produce a higher SNR  in the detectors and are easier to observe compared to binaries with asymmetric masses. Consequently, the relative fraction of observed sources becomes heavily biased in favor of higher mass ratios. 
The detection-weighted mass ratio distribution for CE evolution channel (bottom panel of Fig. \ref{fig:MassRatioCE}) looks very different than the intrinsic one (top panel). The high peak of BH-NS mergers (and unequal mass BH-BH mergers for $f_{\rm mix}=0.5$) has been reduced with respect to more equal mergers. Detection-weighted distribution has a clear, high peak for equal mass systems with $q_{\rm det} \approx 0.9-1.0$ while for intrinsic distribution this peak was barely visible. Figures \ref{fig:q_vs_mtot} for BH-BH mergers and \ref{fig:q_vs_mtot_bhns} for BH-NS mergers help to understand the origin of this peak and the shape of detection weighted distributions. On the top panel of Figure \ref{fig:q_vs_mtot} we present relation between mass ratio $q$ and the average total mass $\overline{m_1+m_2}$ of BH-BH mergers at $z<2.0$ for standard CE development criteria. Plotted results indicates that in case of CE formation channel the more equal BH-BH merger, the more massive it is in average. Average mass of unequal mass ratio BH-BH mergers $q=0.3$ is around $20\msun$. Equal mass BH-BH mergers ($q \in 0.9-1.0$) are significantly more massive, with average mass around $35\msun$. Therefore equal mass mergers more easily detected. Average mass of BH-NS mergers (up to $17 \msun$) for a given mass ratio bin is systematically lower than average mass of BH-BH mergers.
Despite that bias, the total fraction of detectable unequal mass mergers with mass ratio $q_{\rm det}<0.5$ is significant for all SN models within standard CE criteria, constituting $\sim 24\%$ for $f_{\rm mix}=0.5$; $\sim 15 \%$ for $f_{\rm mix}=1.0$; and $\sim 10 \%$ for $f_{\rm mix}=4.0$ (Tab. \ref{tab:mass_ratio_fractions}).
 
\subsubsection{Stable RLOF evolution}

In the case of the revised CE development criteria (stable RLOF BH-BH formation channel, see Fig. \ref{fig:MassRatioPav}) mass ratio distribution of BH-BH mergers is very similar for all tested variants of SN models and both PSN models. In the intrinsic distribution (top panels), the main difference between remnant mass models is a peak of BH-NS mergers, which for models with $f_{\rm mix}=1.0$ and  $f_{\rm mix}=4.0$ is 2-3 times higher and shifted towards more unequal mass ratios ($q \approx 0.1$) comparing to model with $f_{\rm mix}=0.5$ ($q \approx 0.2$). In all tested SN models there is a second broad and high peak for mass ratios in the range $q \in 0.4-0.6$  which dominates the  distribution of BH-BH mergers. This peak and its origin has already been noticed and explained in \cite{Olejak2021a}. In short, the peak is a consequence of overlap of restrict CE development criteria applied for massive donors with masses above $M_{\rm ZAMS}>18 \msun$ (BH progenitors) and the adopted assumptions of rather low orbital angular momentum mass loss during non-conservative RLOF \citep{2008ApJS..174..223B}. In such physical conditions, it is possible to form tight BH-BH system which would merge in Hubble time only if the progenitor system had significantly unequal mass ratio at the onset of the second RLOF. 

The detection-weighted distribution for revised CE development criteria is, as with the the intrinsic distribution, strongly dominated by the peak made of BH-BH mergers with mass ratios $q \in 0.4-0.6$. However, the peak for unequal mass mergers with $q \in 0.1-0.2$, which was present in the intrinsic distribution, nearly disappears in the detection-weighted results with respect to the dominant BH-BH peak. The total fraction of unequal mass mergers with $q_{\rm det}<0.5$ is significant: around $25 \%$ for all remnant mass models. However, for more extreme mass ratio mergers with $q_{\rm det}<0.3$ it quickly becomes negligible, constituting less than 1\%. 

\subsubsection{Mass ratio vs average mass of mergers}

Figures \ref{fig:q_vs_mtot} and \ref{fig:q_vs_mtot_bhns} shows trends in relation between mass ratios and masses of BH-BH and BH-NS mergers respectively for both RLOF stability criteria. On the bottom panel of Figure \ref{fig:q_vs_mtot} we plot the relation between mass ratio $q$ and the average total mass $\overline{m_1+m_2}$ of BH-BH mergers at $z<2.0$ for the revised CE development criteria (stable RLOF formation channel). The trend is much different in the case of BH-BH binaries formed via the CE channel (top panel). The average mass of BH-BH mergers formed via stable RLOF is largest ($\sim 50-60 \msun$) for unequal mass ratios corresponding to a peak $q \in 0.4 - 0.6$ and decreases moving toward equal mass ratio mergers. 

\subsubsection{Comparison with GW detections}

Detection-weighted results for both standard and revised CE development criteria do not fit all of the properties from the current database of LVK detection.  In the  case of the standard CE development criteria, the model with $f_{\rm mix}=0.5$ is able to reconstruct a significant fraction of unequal mass mergers ($q_{\rm det} \leq 0.3$). All three tested models of $f_{\rm mix}$ also produce a good fit to LVK detections for the middle range of mass ratio values: $0.3<q_{\rm det}<0.7$.  For our simulations, we obtain a peak for equal mass ratio systems ($q_{\rm det}>0.9$) while LVK detections show a peak at more unequal mergers: $q_{\rm det} \in 0.7-0.8$. 

In the case of the revised CE development criteria under which most BH-BH mergers formed via stable RLOF, the distribution is dominated by a high, broad peak similar to the LVK detections. However, in our simulations, the peak is shifted towards more unequal mergers, with its center $q_{\rm det}\approx 0.5$ instead of 0.7 (as for LVK). For those models we also do not produce a significant fraction of mergers with $q<0.3$ which is visible in distribution of LVK detections. 

Our results indicate that the distribution of the DCO mass ratio in isolated binary evolution modeling is very sensitive to input physical assumptions. A more extensive parameter study is needed in the future considering several assumptions on e.g. mass and orbital angular momentum loss during RLOF. Such studies could help to verify the model if we are able find an evolution model which fits the LVK observations well with isolated binary evolution.  If not, the LVK observations may require including other formation channels in order to produce all the features in mass ratio distribution of detected DCO mergers.

\subsection{Local merger rate density} \label{subsec: local_densities}

In Table \ref{tab:merger_rates}, we provide the local merger rate densities ($z \approx$ 0) for our three new SN models and all tested physical models (see Sec. \ref{sec: Method}).
The first column is the row number, in the second column we specify used criteria for CE development: standard or revised. The third column stands for the PSN limit used: strong ($\sim 45 \msun$) and revised allowing for formation of more massive BHs. In the fourth column we specify the adopted value of mixing parameter ($f_{\rm mix}=0.5;1.0$ or 4.0) for new remnant mass formula. In columns fifth, sixth and seventh we provide estimated values of the local merger rate densities ($z\approx0$) for BH-BH, BH-NS and NS-NS systems respectively. We also provide a row with the current ranges for DCO merger rates given by recent LVK taking the union of 90\% credible intervals as provided by \cite{LIGOfullO3population2021}. For BH-BH mergers we give two variants of ranges, one with broader range under the assumption of constant rate density versus comoving volume (the upper value) and the second value given for $z = 0.2$ under assumption that BH-BH merger rate evolves with redshift (the lower value).

   \begin{table*} 
     $$ 
         \begin{array}{cccccccc}
            \hline
            \noalign{\smallskip}
            \mbox{No.} & \mbox{CE criteria} & \mbox{PSN limit}  & \mbox{SN model} & \cal{R}_{\mbox{BH-BH}}[\mbox{Gpc$^{-3}$yr$^{-1}$}]  & \cal{R}_{\mbox{BH-NS}}[\mbox{Gpc$^{-3}$yr$^{-1}$}] & \cal{R}_{\mbox{NS-NS}} [\mbox{Gpc$^{-3}$yr$^{-1}$}] \\
            \noalign{\smallskip}
            \hline
            \noalign{\smallskip}
            \mbox{}& &\mbox{GWTC-3} &  & 16-130 & 8-140 & 10-1700\\
            \mbox{}& & ^{z\sim 0.2} &  & ^{18-44} &  & \\
            \hline
            \noalign{\smallskip}
            \mbox{1.}& \mbox{Standard} & \mbox{Revised} & f_{\rm mix}=0.5 & 50 & 10 & 129 \\
            \mbox{2.}& \mbox{Standard} & \mbox{Revised}& f_{\rm mix}=1.0 & 43 & 21 & 119 \\
            \mbox{3.}& \mbox{Standard} & \mbox{Revised} & f_{\rm mix}=4.0 & 46 & 23 & 117 \\
            \noalign{\smallskip}
            \hline
            \noalign{\smallskip}
            \mbox{4.}& \mbox{Standard} & \mbox{Strong} & f_{\rm mix}=0.5 & 61 & 11 & 155 \\
            \mbox{5.}& \mbox{Standard} & \mbox{Strong} & f_{\rm mix}=1.0 & 52 & 26 & 134 \\
            \mbox{6.}& \mbox{Standard} & \mbox{Strong} & f_{\rm mix}=4.0 & 58 & 27 & 143  \\
            \noalign{\smallskip}
            \hline
            \noalign{\smallskip}
            \mbox{7.}& \mbox{Revised} & \mbox{Revised} & f_{\rm mix}=0.5 & 68 & 4 & 124 \\
            \mbox{8.}& \mbox{Revised} & \mbox{Revised} & f_{\rm mix}=1.0 & 75 & 5 & 116 \\
            \mbox{9.}& \mbox{Revised} & \mbox{Revised} & f_{\rm mix}=4.0 & 87 & 6 & 122 \\
            \noalign{\smallskip}
            \hline
            \noalign{\smallskip}
            \mbox{10.}& \mbox{Revised} & \mbox{Strong} & f_{\rm mix}=0.5 & 80 & 5 & 154 \\
            \mbox{11.}& \mbox{Revised} & \mbox{Strong} & f_{\rm mix}=1.0 & 86 & 6 & 138 \\
            \mbox{12.}& \mbox{Revised} & \mbox{Strong}& f_{\rm mix}=4.0 & 102 & 7 & 144 \\
            \hline
            
         \end{array}
         \label{tab:merger_rates}
     $$ 
     \caption{Local merger rate densities ($z\approx0$) for double compact objects (BH-BH, BH-NS and NS-NS) for different tested physical models. In the first row(0.) we give recent (GWTC-3) estimates of LVK \citep{LIGO_Population_2021,LIGO_BHNS_2021}. The columns: CE criteria stands for adopted standard or revised CE development treatment; PSN limit stands for adopted strong ($\sim 45 \msun$) or revised ($\sim 90 \msun$) PSN limit; SN model stands for different adopted $f_{\rm mix}$ parameters (convection growth time); $\cal{R}_{\mbox{DCO}}[\mbox{Gpc$^{-3}$yr$^{-1}$}]$ stands for merger rate densities for different types of double compact object. }
\end{table*}

For all the tested models, the local merger rate densities (at $z\approx0$) for NS-NS systems, which vary from $116$ to $155 \gpy$, are within the wide range constrained by detections of LVK: $10-1700 \gpy$ (the union of 90\% credible intervals). The local merger rate densities for BH-NS systems in our models are rather low and vary from $4$ to $27 \gpy$. Our values are within or close to the lower edge of the LVK range estimates for BH-NS merger rate densities: $8-140 \gpy$ (the union of  90\% credible intervals). BH-BH merger rates for tested models vary from $43$ to $102 \gpy$. It falls into the broader variant of ranges for BH-BH merger rate densities given by LVK: $16-130\gpy$ estimated (as for NS-NS and BH-NS) assuming a constant rate density versus comoving volume and taking the union of 90\% credible intervals \citep{LIGOfullO3population2021}. However, our rates for BH-BH mergers vary from being close to the upper edge to even 2 times the upper edge of the narrow LVK range $18-44 \gpy$, which is the variant accounting for the BH-BH merger rate redshift evolution, estimated at a fiducial redshift ($z = 0.2$).   

\subsubsection{Impact of SN model}

The choice of the adopted SN model most significantly affects merger rates of BH-NS systems in standard CE development criteria. The rate estimated for model with $f_{\rm mix}=0.5$,  $\sim 10 \gpy$, is more than two times lower than the rates estimated for models with $f_{\rm mix}=1.0$ and $f_{\rm mix}=4.0$. Similar effect has been already noticed for previously used {\tt StarTrack} SN models: delay and rapid, and explained by \cite{Drozda2020}. Increased efficiency in the formation of BH-NS mergers for models with more rapid convection growth timescale, as described in more detail in Section \ref{subsec:mass_ratio}, is due to the overlap of the produced mass distribution (particularly the lower mass gap) with the adopted assumption on NS and BH natal kicks. BHs formed via direct collapse in models $f_{\rm mix}=1.0$ and $f_{\rm mix}=4.0$ are more massive and get lower natal kicks than low mass BHs produced in successful SN explosions in model $f_{\rm mix}=0.5$.

Different assumptions on the SN model do not significantly affect BH-BH and NS-NS merger rates neither in the standard nor in the revised CE development criteria. In the case of NS-NS, the difference between results for tested remnant models is up to $\sim 10 \%$ while for BH-BH mergers up to $\sim 25 \%$. 

\subsubsection{Impact of assumption on PSN}

The rates for all types of DCO mergers are slightly affected by the adopted PSN model (strong or revised) which changes them by around $\sim 20 \%$ due to different normalisation caused by different IMF range (see Sec. \ref{sec: Method}). For the revised PSN model, we extend the range for the initial mass of the stars to $200 \msun$. This change makes it possible to create heavy BH-BH mergers with component masses $m_1+m_2\geq 100\msun$. Such heavy BH-BH mergers are, however, rare (see Figures \ref{fig:MassDistribution_BHBH_CE} and \ref{fig:MassDistribution_BHBH_Pav}, Sec. \ref{subsec: mass_distribution}) and do not constitute quantitatively significant fraction of all BH-BH mergers. Production of massive stars with initial masses $150-200 \msun$ takes some part of the total stellar mass (constant for all models) and slightly reduces the rates of DCO mergers in the revised PSN model.

\subsubsection{Impact of RLOF stability criteria}

Among all the tested models in this study factors, the merger rates of BH-BH and BH-NS are mostly affected by the used CE development criteria. As shown in \cite{Olejak2021a}, different assumption on RLOF stability may lead to different dominant formation scenario for DCO mergers. In the revised CE development criteria BH-BH and BH-NS progenitor systems are less likely to initiate a CE phase; for example, massive donors with $M_{\rm ZAMS}>18\msun$ (BH progenitors) RLOF is assumed to be stable for much wider parameter space than in the standard criteria. The rates of BH-BH mergers between the revised and the standard CE development criteria vary up to a factor of $\sim 2$ while BH-NS even by a factor of $\sim 5$. However, due to a coincidence of the peak in star formation at $z\approx 2.0$ with the length of the BH-BH time delays\footnote{Time delay is the time since formation of double compact object system (NS-NS, BH-NS, BH-BH) till its merger due to gravitational waves emission.}, which for stable RLOF phases are on average much longer than for CE formation channel, we obtain a non-intiutive result for BH-BH merger rates in the local Universe ($z\approx 0$). As shown in Figure \ref{fig:BHBH_rates_redshift} (Appendix \ref{sec:redshift_BHBH_rates}), for most of the redshifts ($z\geq0.5$) the BH-BH merger rate density is systematically a few times lower for the revised criteria dominated by stable RLOF formation channel than for the standard criteria dominated by CE formation channel. However, due to longer BH-BH time delays for the stable RLOF channel, the peak in the merger rates related to increased star formation (at $z\approx 2.0$) is shifted towards lower redshifts. As a result for the lowest redshifts ($z\leq0.5$) BH-BH merger rate density is $\sim 2$ times higher for the revised CE development criteria than for the standard one.  

As discussed in Section \ref{sec: Discussion}, the physical parameter space tested within this study (12 models) is very limited and there are several other uncertain parameters and unconstrained processes which may significantly influence local merger rate densities. In this work, we focus on the impact of the adopted new remnant mass formulas. A few examples of works which provide estimates of DCO merger rates studying influence of other physical factors are \cite{2013ApJ...779...72D,Neijssel2019}or\citep{Broekgaarden2021}.  

\section{Discussion of caveats and uncertainties} \label{sec: Discussion}

At this point we would like to mention some of the uncertainties in the calculations made in this work. In this study we made a lot of assumptions related to the physical modeling of single and binary stellar evolution. Besides the stellar collapse and SN engine mechanism which is the main subject of this work, there are many other uncertain processes that dictate the formation of DCO mergers. Some examples of such processes or parameters which may significantly influence physical properties of DCO mergers are: the metallicity-specific star formation rate density and initial mass function \citep{2020A&A...636A..10C,Santoliquido2021,2021arXiv210302608B,Briel2021,Kroupa2021},
stellar winds \citep{Vink2001, Sander2022}, convection and overshooting \citep{Klencki2020, Belczynski2022}, NS and BH natal kicks 
\citep{mandel2021,Igoshev2021}, type and rate of mass transfer and the outcome of stable/unstable RLOF \citep{Vinciguerra2020,Howitt2020,Bavera2020,Olejak2021a,VignaGomez2022}. The assumptions adopted for this work are described in Section \ref{sec: Method}. Here we give examples of few uncertainties closely related to the new adopted formulas for SN remnant masses.

The {\tt StarTrack} population synthesis code uses analytic fits to the results of detailed evolution models given by \cite{Hurley2000} and \cite{Hurley_2002}. It has various possible consequences like a limited tested parameter space for detailed stellar physics or possible numerical artifacts. A specific example of possible over- or under-estimation of the stellar structure parameters may be due to stopping the simulations before the star reaches collapse. The fits provided by \cite{Hurley2000} are based on results which tracked nuclear evolution of individual star and its structure parameters such as the core mass only till the end its core helium burning phase. However, the mass of the helium or carbon-oxygen core, the parameters which are used in this study e.g. in the new remnant mass formulas or in PSN modeling, still grow in the later part of the stellar evolution. 

Using the carbon-oxygen core $M_{\rm CO}$ as a main tracer to estimate the fate of stellar collapse and the final remnant mass is not a perfect solution also for other reason. The value of $M_{\rm CO}$ does not capture the full structure of the pre-SN star nor the important features like the density gradient in the silicon and oxygen shells surrounding the iron core. Known in the literature as the compactness parameter, being a function of the density gradient, could be useful in making more precise predictions for remnant masses. Studies, e.g. \cite{Sukhbold2014}, indicate that compactness parameter may be a highly non-monotonic function of the ZAMS mass of the pre-SN star.
Unfortunately, most of the population synthesis codes including {\tt StarTrack} do not have an access to the detailed properties of stellar cores needed to estimate compactness parameter. In addition, it is known that, although the compactness parameter is a reasonable guide to the fate of the star, it also has its limitations~\citep{2001ApJ...550L.169A,2016ApJ...818..124E, 2020MNRAS.491.2715B,Fryer2022}. On the other hand, using fits to available results for the relation between ZAMS mass of pre-SN star and the final compactness parameter derived using detailed codes for single stellar evolution would not be a good approach as it would neglect the impact of mass transfer episodes (CE or stable RLOF), which are common and crucial for DCO mergers progenitors. Therefore, in this study we decided to use $M_{\rm CO}$ as a main parameter which seems to be the best compromise despite the caveats. 

Another uncertainty is the M$_{\rm crit}$ parameter in formula \ref{eq:mrem} which in this work is set to 5.75 $\msun$ and stands for the critical mass for the switch from NS to BH formation (threshold on mass of carbon-oxygen core). This is a parameter which is not yet constrained by observations. Different assumptions on M$_{\rm crit}$ impacts final remnant mass distribution, e.g. the width of the lower mass gap. See Figure \ref{fig:MassGap_mcrit4.75}, where for comparison, we provide same results as in Figure \ref{fig:MassGap} but for $M_{\rm crit}=4.75M_\odot$.

Finally, we point out that the distribution of remnant masses could look very different for rapidly rotating progenitors. If the angular momentum of the collapsing star is large enough, the SN explosion would be driven by a magnetar engine or NS accretion disk jet instead (See Sec. 4 of \cite{Fryer2022}). Rapidly-rotating BH progenitors are believed to form accretion disks at the collapse and possibly drive an energetic jet. The jet could eject significant fraction of the stellar envelope and be responsible for long gamma ray bursts phenomenon \citep{Woosley1993,MacFadyen2001}. Such an additional ejection of mass would lead to formation of less massive compact objects (NSs and BHs) compared to remnant masses estimated by the new formulas applied for this study.  However, as shown by \cite{Fryer2022} (see Fig. 11), such rapid rotators are not expected to constitute significant fraction of BH progenitors as the typical total angular momentum of pre-SN stars is usually expected to be way below the required limit. Also under-resolved simulations of magnetic field as well as the eventual jet production is not yet predictive.    

\section{Conclusions} \label{sec: Conclusions}

In this study we employ {\tt StarTrack} population synthesis code to test how new formulas for SN models by \cite{Fryer2022} affect DCO mergers formed via isolated binary evolution. New formulas for remnant masses allow to probe the convection growth timescales in wide spectrum in contrast to previously used extreme cases of delayed and rapid SN models \citep{Fryer2012}. The different assumption for the convection growth timescale impacts the depth and width of the lower mass gap which may exist between maximum possible mass of NS and minimum mass of BH. In addition to new models of SN, we test two variants of PSN: strong PPSN/PSN model which limits BH mass to $\sim 45 \msun$ and revised which shifts the limit to higher BH masses (PSN for helium cores: $90\msun <M_{\rm He} <175 \msun$). We also present results for two CE development criteria: the standard model with BH-BH mergers formation mainly via CE and the revised with BH-BH mergers formation mainly via stable RLOF. \\
A summary of the results in this study is: \\
i). Different assumptions of the convection growth timescale have a strong impact on the width and depth of the lower mass gap in remnant mass distribution for both single and binary evolution. The most-rapid SN model tested, the variant with $f_{\rm mix}=4.0$ (rapid growth of the convection $<$10ms that then develops an explosion in the first 100ms) produces wide and deep mass gap between $\sim 2-7 \msun$. In contrast, for the most-delayed SN model tested with $f_{\rm mix}=0.5$ (growth time closer to 100ms where the explosion can take up to 1s) the fraction of compact objects within the lower mass gap range is 1-2 orders of magnitude larger than in the model with $f_{\rm mix}=4.0$. Models with mixing parameter $f_{\rm mix} \approx 2.5$ produce a gap between $\sim 2-5 \msun$ which is in good agreement with observations.\\
ii). The mass distribution of DCO mergers: mass of the primary $m_1$, secondary $m_2$ and their sum, is sensitive to adopted SN model up to total merger mass of $m_1+m_2 \lesssim 35\msun$.\\ 
iii). The SN model with $f_{\rm mix}=0.5$ produces significant fraction of massive NSs with mass $\in 1.5-2.5 \msun$ which may be important in studying the origin of massive NS systems like GW190425 \citep{massive_NSNS2020}. \\
iv). The choice of SN model affects significantly the intrinsic mass ratio distributions of BH-BH mergers but only in the CE formation scenario. In the intrinsic mass ratio distribution of BH-BH mergers (CE formation) with adopted $f_{\rm mix}=0.5$ there is a broad peak for unequal mass mergers $q \in 0.1-0.3$, while BH-BH mass ratio distribution for $f_{\rm mix}=4.0$ has two peaks at $q \in 0.5-0.7$ and for equal mass ratio systems $q \approx 1.0$ instead. Mass ratio distributions of BH-BH mergers formed mainly through stable RLOF is dominated by a broad, high peak for $q \in 0.4-0.6$ independently on SN model choice. 
\\
v). The choice of SN model has the most significant effect on local merger rates densities for BH-NS mergers. Rates of BH-NS mergers may even vary by a factor of $\sim 3$ between two extreme cases  $f_{\rm mix}={4.0}$ and $f_{\rm mix}={0.5}$. Such a difference is due to an overlap of the remnant mass distribution and fraction of low mass compact objects with our assumption of natal kicks (inversely proportional to the mass of compact object). The level of discrepancy between rates for different SN models is sensitive to other tested physical assumptions. Due to wide mass gap which means the dearth of massive NSs or low mass BHs in model with $f_{\rm mix}=4.0$ there is a high peak for very unequal mass ratio BH-NS mergers with $q \sim 0.1$.  \\
vi). Different assumptions on PSN limit do not significantly influence the mass ratio distribution of BH-NS and BH-BH mergers. Extending the PSN limit to higher masses in revised PSN model allows to have a low tail of more massive mergers with total mass $m_1 + m_2 > 90 \msun$. For strong PPSN/PSN model the total mass of DCO mergers reaches up to  $m_1 + m_2 \approx 90 \msun$, at which the value of the total merger mass has a slight peak due to partial mass ejection in PPSN. \\
vi). The adopted CE development criteria differ only for BH progenitors and therefore, the properties of NS-NS binaries are not affected by used RLOF stability criteria. The choice of these criteria influence mainly properties of BH-BH mergers while BH-NS mergers are partly affected. \\
vii). Our simulations for binary systems, similar to results for single evolution presented in \cite{Fryer2022}, show that eventual stochasticity in stellar structure should not affect statistical picture for large probe of compact object population. \\
The SN models which partially fill the lower mass gap seem promising as such predictions could match current EM observational results on the suppressed number of compact objects in the lower mass gap. In this work, we refrain from making any strong conclusions about the convection growth timescale due to many uncertainties (see some caveats Sec. \ref{sec: Discussion}). We also still do not know what fraction of LVK mergers may come from different evolutionary channels (not necessarily isolated binary evolution). However, further studies constraining detailed theoretical models by observations could bring progress toward understanding the evolution of massive stars (single and binary) and also the formation channels of detected DCO mergers. In addition, the number of future DCO detections is predicted to increase dramatically with next generation detectors \citep[e.g.][]{Borhanian2022}. A much larger data base of parameters of DCO mergers will help to put more constrains on SN mechanism and the origin of detected signals.   

\section*{Acknowledgements}
K.B. and A.O. acknowledge support from the Polish National Sci-
ence Center (NCN) grant Maestro (2018/30/A/ST9/00050). A.O. is also supported by the Foundation for Polish Science (FNP). The work by CLF was supported by the US Department of Energy through the Los Alamos National Laboratory. Los Alamos National Laboratory is operated by Triad National Security, LLC, for the National Nuclear Security Administration of U.S.\ Department of Energy (Contract No.\ 89233218CNA000001).

\section*{DATA AVAILABILITY}
Data available on request. The results of simulations underlying this article will be shared on a request sent to the corresponding author: aolejak@camk.edu.pl.




\bibliographystyle{mnras}
\bibliography{ms} 




\appendix

\section{Critical mass for a BH formation vs. lower mass gap} \label{sec:Mcrit}

\begin{figure*}
	\includegraphics[width=12 cm]{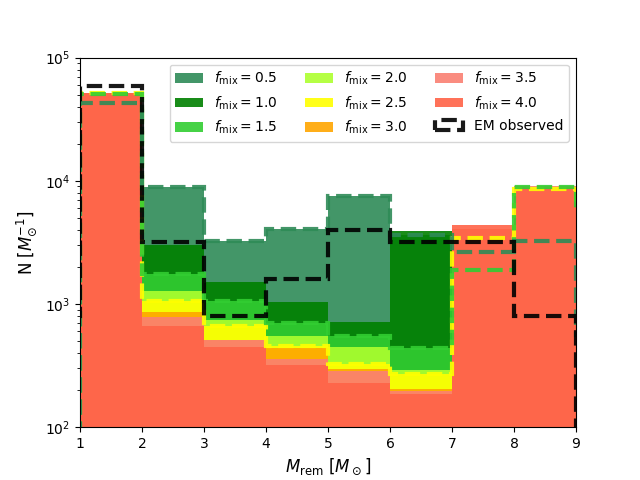}
    \caption{Histogram (binsize $1\msun$) of the remnant masses of the probe of $10^5$ single stars for adopted initial mass function for eight new SN model prescriptions with $f_{\rm mix}$ in a range 0.5-4.0 and $Z=1.0 Z_{\odot}$. Results for different critical mass of carbon oxygen core for a BH formation, $M_{\rm crit}=4.75M_\odot$ (see Eq. \ref{eq:mrem}) than adopted for this study $M_{\rm crit}=5.75M_\odot$ (see Fig. \ref{fig:MassGap}).} 
    \label{fig:MassGap_mcrit4.75}
\end{figure*}

\section{Redshift evolution of merger rate density of BH-BH mergers} \label{sec:redshift_BHBH_rates}

\begin{figure*}
	\includegraphics[width=12 cm]{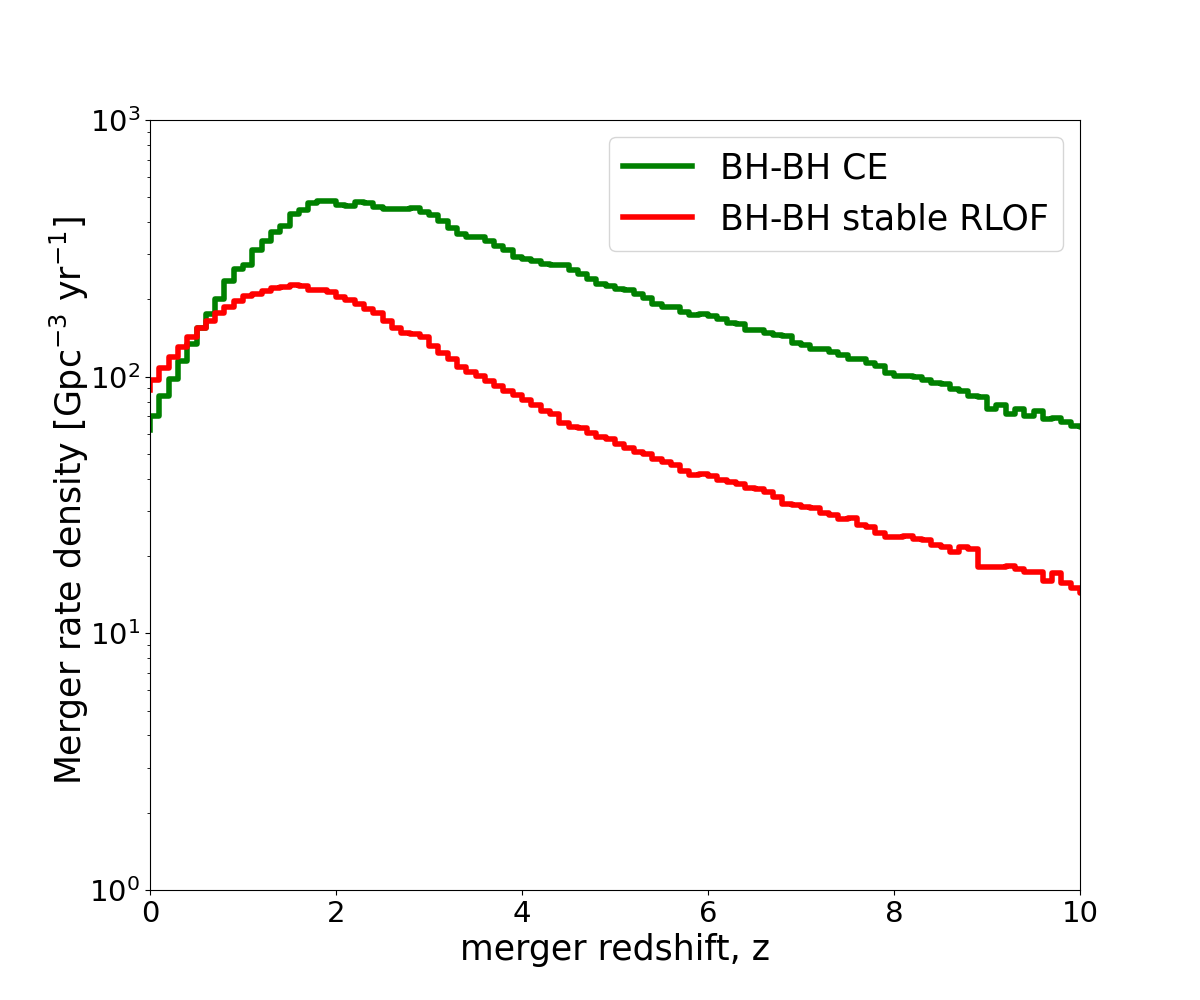}
    \caption{BH-BH merger rate density evolution with redshift $z$. Comparison between standard and revised CE development criteria. With the green line we plot result for a model with the standard criteria for which BH-BH forms mainly through CE phase. The results for the revised criteria for which BH-BH forms mainly through stable RLOF is plotted with a red line. Note that, for most of the redshifts, the merger rate density for the revised criteria (BH-BH stable RLOF) is systematically a few times lower (up to one order of magnitude) than for the standard one. However, due to long time delayes of BH-BH mergers in stable RLOF channel, the peak in merger rates related to a peak in star formation (at $z\approx 2.0$) is shifted to the left (towards lower redshifts). This leads to non-intuitive result that for the local Universe $z \approx 0$ BH-BH merger rate density is higher for stable RLOF than for CE formation channel. }
    \label{fig:BHBH_rates_redshift}
\end{figure*}

\section{Stochasticity} \label{sec:Stoachsticity}

Here we check how assumption on some stochasticity in the in pre-SN stellar structure, reported by several studies \citep{Patton2020,Laplace2021,Schneider2021} may influence the mass distribution of cosmological DCO mergers in a similar way as \cite{Fryer2022} did for single star evolution. We mimic the stochasticity effect by including in our synthetic population a fraction of stars $f_{\rm stoch}$ which may have much different mass of their carbon-oxygen core $M_{\rm CO}$ at pre-SN evolution stage comparing to the value predicted by standard {\tt StarTrack} calculations. For this fraction of stars we generate value of $M_{\rm CO}$ from the flat distribution in range between 10\% of original value $M_{\rm CO}$ and the total pre-SN mass of the star, so:

\begin{equation} 
M_{\rm CO}^{\rm stoch} = {\rm rand}(0.1 M_{\rm CO},M_{\rm fin})    
\label{eq:stochasticity}
\end{equation}

\begin{figure}
    	\includegraphics[width=8.5 cm]{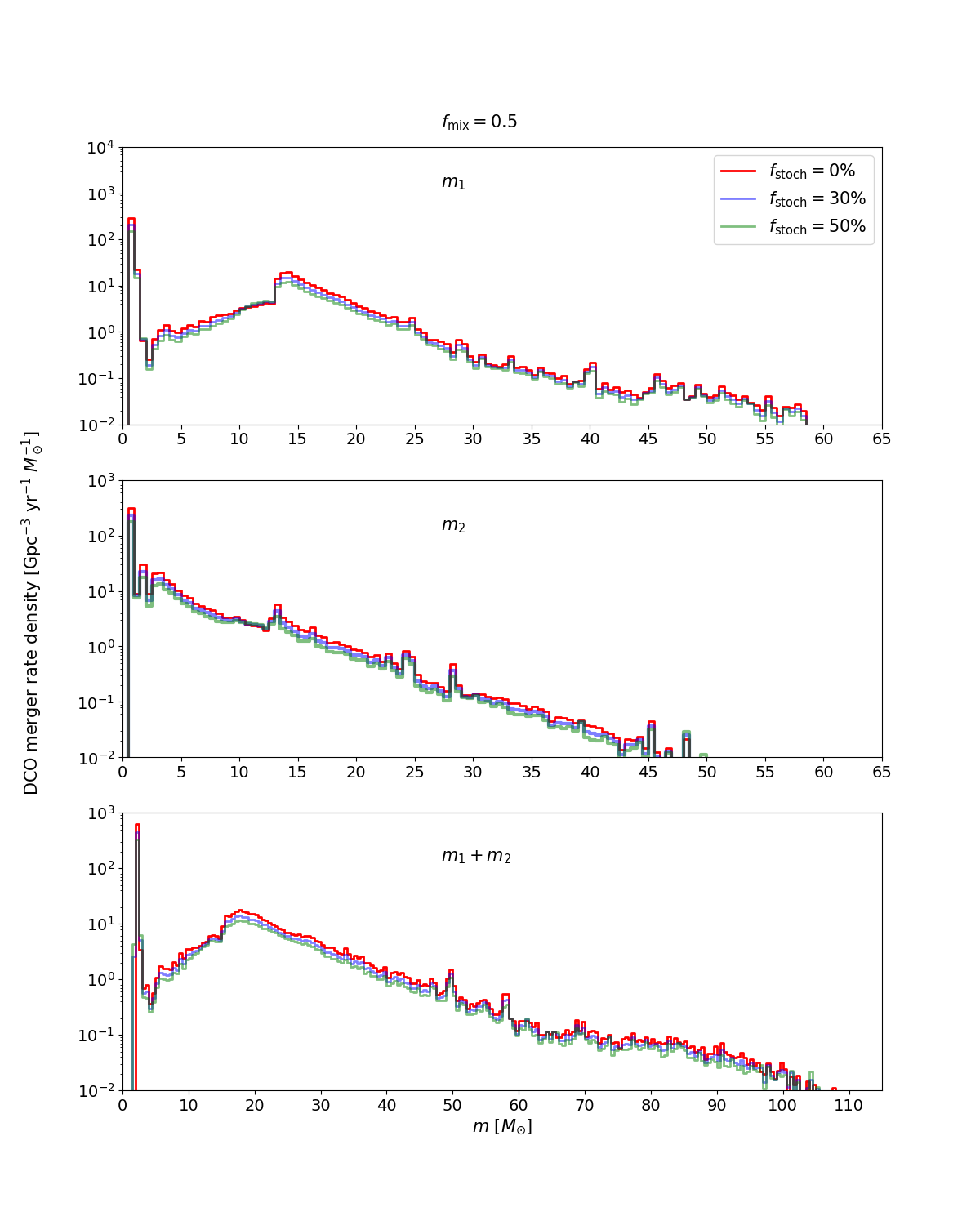}
    \caption{Mass distribution of DCO merger masses ($z<1.0$) with different fractions of compact object progenitors with stochastic stellar structure (Eq. \ref{eq:stochasticity}): 0\% (red line), 30\% (blue line) and 50\% (green line). Results for new remnant mass formula with f$_{\rm mix}=0.5$, a standard CE development criteria and revised PSN limit (see Sec. \ref{sec: Method}). The top panel - distribution of primary mass $m_1$, the middle panel - secondary mass $m_2$, the bottom panel - $m_1+m_2$.}
    \label{fig:Stoch0.5}
\end{figure}

\begin{figure}
	\includegraphics[width=8.5 cm]{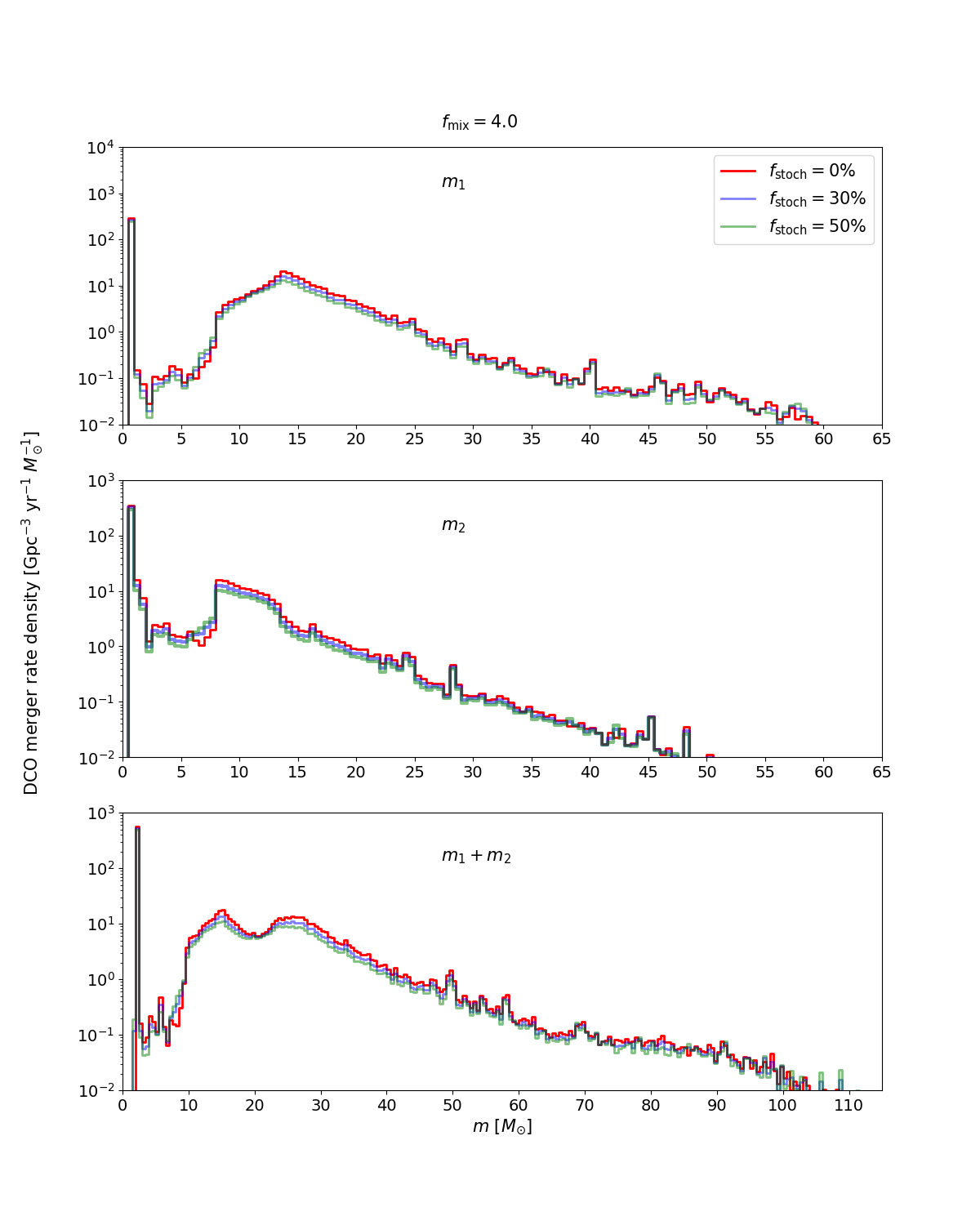}
    \caption{Mass distribution of DCO merger masses ($z<1.0$) with different fractions of compact object progenitors with stochastic stellar structure (Eq. \ref{eq:stochasticity}): 0\% (red line), 30\% (blue line) and 50\% (green line). Results for new remnant mass formula with f$_{\rm mix}=4.0$, a standard CE development criteria and revised PSN limit (see Sec. \ref{sec: Method}). The top panel - distribution of primary mass $m_1$, the middle panel - secondary mass $m_2$, the bottom panel - $m_1+m_2$.}
    \label{fig:Stoch4.0}
\end{figure}

The results are shown in Figures \ref{fig:Stoch0.5} and \ref{fig:Stoch4.0} where we plot mass distribution of DCO mergers for two examples of new SN models:  f$_{\rm mix}=0.5$ and f$_{\rm mix}=4.0$ respectively. Figures show results for three variants of included fractions of stochastic-structure progenitors (Eq. \ref{eq:stochasticity}). Those fractions are 0\% (red line), 30\% (blue line) and 50\% (green line). On the top panel we show distribution of primary mass $m_1$, on the middle panel secondary mass $m_2$ and on the bottom panel the sum $m_1+m_2$. The plotted results are for a physical model with a standard CE development criteria and revised PSN limit (see Sec. \ref{sec: Method}).

\section{StarTrack vs MOBSE new remnant mass prescriptions} \label{sec:MOBSEcomparios}

In this section we make a brief comparison of outcomes for our new SN prescriptions parameterizing convection growth timescale with other recent prescription for stellar remnants parameterizing the fraction of the hydrogen envelope that is accreted by the BH during collapse \citep{Dabrowny2021}. The formula by \cite{Dabrowny2021} tested with MOBSE population synthesis code \citep{Giacobbo2018} use compactness parameter $\zeta_{2.5}$ approximated with the formula $$\zeta_{2.5} \approx 0.55 - 1.1 (M_{\rm CO}/1 \msun)^{-1}$$ as a threshold between NS and BH formation. The value used in their simulations $\zeta_{2.5} = 0.365$ was calibrated to match results for previous rapid SN model by \citep{Fryer2012}. The mass of the BH is calculated with the formula:
$$M_{\rm BH} = M_{\rm He}+ f_{\rm H}(M-M_{\rm He})$$

Where $M_{\rm He}$ is the mass of the helium core, $M$ - pre-SN mass of the star while $f_{\rm H}$ is the parameter which determines assumed fraction of mass of the hydrogen envelope accreted on the BH during collapse.
For the main tested model in their studies this fraction was $f_{\rm H}=0.9$.

\begin{figure}
	\includegraphics[width=8.5 cm]{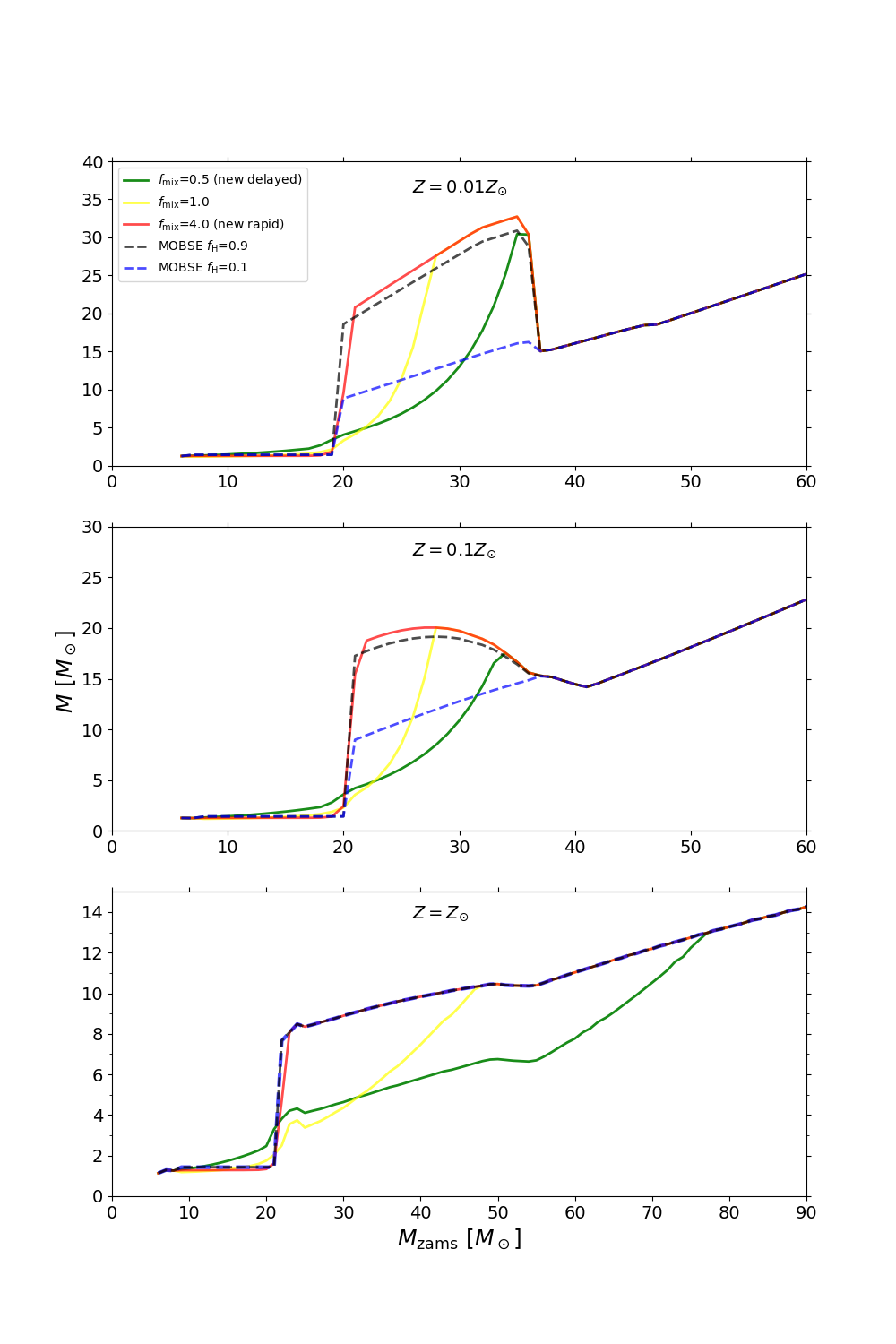}
    \caption{Relation between initial mass $M_{\rm ZAMS}$ of progenitor star and the final remnant mass for three new SN models \citep{Fryer2022} parameterizing convection growth timescale: $f_{mix}=0.5$, $f_{mix}=1.0$ and $f_{mix}=4.0$ and two remnant mass variants parameterizing the fraction of the hydrogen envelope \citep{Dabrowny2021}:$f_{\rm H}=0.9$ and $f_{\rm H}=0.1$.}
    \label{fig:SN_comparison}
\end{figure}

In Figure \ref{fig:SN_comparison} we show relation between the initial mass of the progenitor star and its final remnant mass for three cases of our new SN models and two variants of \cite{Dabrowny2021} formulas. 

As the threshold for BH formation was calibrated to rapid SN model in \citep{Fryer2012} and \cite{Dabrowny2021} prescription with $f_{\rm H}=0.9$ assumes almost direct collapse to BH, as expected, the outcome for the model is almost the same as our new rapid SN model with $f_{\rm mix}=4.0$. We also test other extreme assumption with very low fraction of accreted hydrogen envelope $f_{\rm H}=0.1$. If the final pre-SN star still has hydrogen envelope, this model produces systematically lower BH masses comparing to our model with $f_{\rm mix}=4.0$ and \cite{Dabrowny2021} with $f_{\rm H}=0.9$. However, as the minimum mass of a BH in \cite{Dabrowny2021} formulas is equal to its pre-SN helium core, this model produce a mass gap even for low fraction of accreted envelope $f_{\rm H}=0.1$ (there is steep increase between the masses of NSs and lowest masses of BHs, see Fig. \ref{fig:SN_comparison}). Therefore \cite{Dabrowny2021} parameterization would not allow for a smooth transition between having a deep mass gap and a remnant mass distribution filled with significant number of $2-5 \msun$ compact objects as in case of our \cite{Fryer2022} models with wide spectrum of convection growth timescales. 

\bsp	
\label{lastpage}
\end{document}